\RequirePackage{fix-cm} % Fix LaTeX2e bugs.
\documentclass[a4paper, twoside, reqno, dvips, 12pt]{amsart}
\usepackage{fixltx2e}   % Fix LaTeX2e bugs.

\usepackage{etex}

\usepackage[latin1]{inputenc}
\usepackage[T1]{fontenc}

\usepackage{esint}
\usepackage{dsfont}
\usepackage{xspace}
\usepackage{amsgen}
\usepackage{amsthm}
\usepackage{amssymb}
\usepackage{amsmath}
\usepackage{wasysym}
\usepackage{upgreek}
\usepackage{amsfonts}
\usepackage{stmaryrd}
\usepackage{scalerel}
\usepackage{mathtools}
\usepackage{stackengine}

\usepackage{relsize}
\usepackage{textcomp}
\usepackage[mathcal, mathscr]{euscript}

\usepackage{mathrsfs}
\DeclareMathAlphabet{\mathscrbf}{OMS}{mdugm}{b}{n}

\usepackage{a4wide}

\usepackage[dvipsnames, table]{xcolor}
\definecolor{bckg}{RGB}{20.8, 20.8, 20.8}
\definecolor{oneblue}{rgb}{0.0, 0.0, 0.85}
\definecolor{Lightblue}{RGB}{214, 214, 214}
\definecolor{bluepigment}{rgb}{0.2, 0.2, 0.6}
\definecolor{charcoal}{rgb}{0.21, 0.27, 0.31}
\definecolor{denimblue}{rgb}{0.08, 0.38, 0.74}
\definecolor{Lightgray}{rgb}{0.89, 0.89, 0.89}
\definecolor{darkgrey}{rgb}{0.273, 0.281, 0.30}
\definecolor{darkelectricblue}{rgb}{0.33, 0.41, 0.47}

\usepackage[sort&compress, comma, square, numbers]{natbib}

\usepackage{psfrag}
\usepackage{graphicx}
\usepackage{adjustbox}
\usepackage[tight]{subfigure}
\usepackage{morefloats}
\usepackage{indentfirst}

\usepackage[usenames, dvipsnames, pdf]{pstricks}
\usepackage{epsfig}
\usepackage{pst-grad} % For gradients
\usepackage{pst-plot} % For axes

\usepackage{rotating}
\usepackage{pdflscape}

\usepackage{acronym}
\usepackage{microtype}
\usepackage[labelsep=period,%
            labelfont={bf,sf,color=bluepigment},%
            justification=raggedright]{caption}

\usepackage[perpage, symbol]{footmisc}

\usepackage[colorlinks,
           urlcolor=oneblue,
           linkcolor=denimblue,
           citecolor=NavyBlue,
           bookmarksopen=false,
           pdfpagemode=UseNone,
           pagebackref]{hyperref}

\usepackage[explicit]{titlesec}

\titleformat{\section}[block]
  {\color{NavyBlue}\Large\sffamily\bfseries}
  {}
  {0.0em}
  {\colorbox{bckg!5}{\strut\parbox{\dimexpr\linewidth-2\fboxsep\relax}{\thesection. #1}}}
  [\vspace*{0.33em}]

\titleformat{name=\section,numberless}[block]
  {\color{NavyBlue}\Large\sffamily\bfseries}
  {}
  {0.0em}
  {\colorbox{bckg!5}{\strut\parbox{\dimexpr\linewidth-2\fboxsep\relax}{#1}}}
  [\vspace*{0.33em}]

\titleformat{\subsection}
  {\color{NavyBlue}\large\sffamily\bfseries}
  {}
  {0.0em}
  {\colorbox{bckg!5}{\parbox{\dimexpr\linewidth-2\fboxsep\relax}{\thesubsection. #1}}}
  [\vspace*{0.33em}]

\titleformat{name=\subsection,numberless}
  {\color{NavyBlue}\Large\sffamily\bfseries}
  {}
  {0em}
  {\colorbox{bckg!5}{\parbox{\dimexpr\linewidth-2\fboxsep\relax}{#1}}}
  [\vspace*{0.33em}]

\titleformat{\subsubsection}
  {\color{bluepigment}\sffamily\normalsize\bfseries}
  {\thesubsubsection}
  {0.5em}
  {#1}
  [\vspace*{0.33em}]

\titleformat{\paragraph}[runin]
  {\color{bluepigment}\sffamily\small\bfseries}
  {}
  {0em}
  {#1}

\titlespacing{\section}{0.0em}{1.5em plus 2pt minus 2pt}%
{1.0em plus 2pt minus 2pt}[0em]
\titlespacing{\subsection}{0.5em}{1.5em plus 2pt minus 2pt}%
{1.0em}[0em]
\titlespacing{\subsubsection}{0.5em}{1.5em plus 2pt minus 2pt}%
{1.0em plus 2pt minus 2pt}[0em]

\usepackage{titletoc}

\contentsmargin{0.5em}
\newlength{\tocsep} 
\titlecontents{section}[\tocsep]
  {\addvspace{10pt}\bfseries\sffamily}
  {\contentslabel[\thecontentslabel]{\tocsep}}
  {}
  {\ \titlerule*[0.75pc]{.}\ \thecontentspage}
  []
\titlecontents{subsection}[\tocsep]
  {\addvspace{8pt}\sffamily}
  {\contentslabel[\thecontentslabel]{\tocsep}}
  {}
  {\ \titlerule*[0.5pc]{.}\ \thecontentspage}
  []
\titlecontents*{subsubsection}[\tocsep]
  {\addvspace{2pt}\footnotesize\sffamily}
  {}
  {}
  {\ \titlerule*[0.35pc]{.}\ \thecontentspage}
  [\\*]

\makeatletter
\def\@setauthors{%
  \begingroup
  \def\thanks{\protect\thanks@warning}%
  \trivlist
  \centering\footnotesize \@topsep30\p@\relax
  \advance\@topsep by -\baselineskip
  \item\relax
  \author@andify\authors
  \def\\{\protect\linebreak}%
  \textsc{\normalsize\textcolor{darkelectricblue}{\authors}}%
  \ifx\@empty\contribs
  \else
    ,\penalty-3 \space \@setcontribs
    \@closetoccontribs
  \fi
  \endtrivlist
  \endgroup
}
\def\@settitle{\begin{center}%
  \baselineskip14\p@\relax
    \bfseries
    \textsc{\Large\textcolor{charcoal}{\@title}}
  \end{center}%
}
\makeatother

\usepackage{enumitem}
\setlist[description]{%
  topsep=30pt,               % space before start / after end of list
  itemsep=5pt,               % space between items
  font={\bfseries\sffamily\color{NavyBlue}}, % if colour is needed
}

\usepackage{fancyhdr}
\usepackage{lastpage}

\newcommand*\Title{\textcolor{bluepigment}{Turbidity flow modelling}}
\newcommand*\Authors{\textcolor{bluepigment}{V.~Liapidevskii, D.~Dutykh \& M.~Gisclon}}
\newcommand*{\plogo}{\textcolor{gray}{{\texttt{arXiv.org} / \textsc{hal}}}} % Generic publisher logo

\numberwithin{equation}{section}

\newtheorem{remark}{Remark}

\newcommand{\m}{{\sf m}}
\newcommand{\s}{{\sf s}}
\newcommand{\cm}{{\sf cm}}

\newcommand{\R}{\mathds{R}}

\newcommand{\ud}{\mathrm{d}}

\newcommand{\E}{\mathcal{E}}
\newcommand{\M}{\mathcal{M}}
\renewcommand{\phi}{\varphi}
\renewcommand{\beta}{\upbeta}
\renewcommand{\a}{\mathcal{A}}

\renewcommand{\leq}{\leqslant}
\renewcommand{\geq}{\geqslant}

\newcommand{\rhob}{\bar{\rho}}
\renewcommand{\P}{\mathcal{P}}
\renewcommand{\alpha}{\upalpha}
\newcommand{\x}{\boldsymbol{x}}
\renewcommand{\theta}{\vartheta}
\newcommand{\vO}{\boldsymbol{0}}
\renewcommand{\b}{\boldsymbol{b}}
\renewcommand{\u}{\boldsymbol{u}}
\newcommand{\const}{\mathrm{const}}

\newcommand{\F}{\mathscr{F}}
\newcommand{\Aa}{\mathscr{A}}
\newcommand{\Bb}{\mathscr{B}}
\newcommand{\Dd}{\mathscr{D}}
\newcommand{\Jj}{\mathscr{J}}

\newcommand{\Fr}{\mathrm{Fr}}

\newcommand{\cf}{\emph{c.f.\xspace}}
\newcommand{\ie}{\emph{i.e.\xspace}}
\newcommand{\eg}{\emph{e.g.\xspace}}
\newcommand{\etc}{\emph{etc.\xspace}}

\renewcommand{\sim}{\thicksim}

\newcommand{\abs}[1]{\lvert\, #1\, \rvert}

\newcommand{\pd}[2]{\frac{\partial #1}{\partial\/ #2}}
\newcommand{\od}[2]{\frac{\mathrm{d} #1}{\mathrm{d}\/#2}}

\newcommand{\eqdef}{\mathop{\stackrel{\,\mathrm{def}}{:=}\,}}
\newcommand{\defeq}{\mathop{\stackrel{\,\mathrm{def}}{=:}\,}}

\newcommand{\half}{{\textstyle{1\over2}}}
\newcommand{\third}{{\textstyle{1\over3}}}

\newcommand{\fourninth}{{\textstyle{4\over9}}}

\usepackage{acronym}
\acrodef{bvp}[BVP]{Boundary Value Problem}
\acrodef{NSWE}{Nonlinear Shallow Water Equations}

\begin{document}

\title[\Title]{On the modelling of shallow turbidity flows}

\author[V.~Liapidevskii]{Valery~Yu.~Liapidevskii}
\address{\textbf{V.~Liapidevskii:} Novosibirsk State University and Lavrentyev Institute of Hydrodynamics, Siberian Branch of RAS, 15 Av.~Lavrentyev, 630090 Novosibirsk, Russia}
\email{vliapid@mail.ru}
\urladdr{https://www.researchgate.net/profile/V\_Liapidevskii/}

\author[D.~Dutykh]{Denys Dutykh$^*$}
\address{\textbf{D.~Dutykh:} LAMA, UMR 5127 CNRS, Universit\'e Savoie Mont Blanc, Campus Scientifique, F-73376 Le Bourget-du-Lac Cedex, France and Univ. Grenoble Alpes, Univ. Savoie Mont Blanc, CNRS, LAMA, 73000 Chamb\'ery, France}
\email{Denys.Dutykh@univ-smb.fr}
\urladdr{http://www.denys-dutykh.com/}
\thanks{$^*$ Corresponding author}

\author[M.~Gisclon]{Marguerite Gisclon}
\address{\textbf{M.~Gisclon:} Univ. Grenoble Alpes, Univ. Savoie Mont Blanc, CNRS, LAMA, 73000 Chamb\'ery, France}
\email{Marguerite.Gisclon@univ-smb.fr}
\urladdr{http://www.lama.univ-savoie.fr/~gisclon/}


%%% ----------------------------------------------------------------------- %%%

\begin{titlepage}
\thispagestyle{empty} % Remove page numbering on this page
\noindent
{\Large Valery \textsc{Liapidevskii}}\\
{\it\textcolor{gray}{Lavrentyev Institute of Hydrodynamics, Novosibirsk, Russia}}\\
{\it\textcolor{gray}{Novosibirsk State University, Novosibirsk, Russia}}
\\[0.02\textheight]
{\Large Denys \textsc{Dutykh}}\\
{\it\textcolor{gray}{CNRS--LAMA, Universit\'e Savoie Mont Blanc, France}}
\\[0.02\textheight]
{\Large Marguerite \textsc{Gisclon}}\\
{\it\textcolor{gray}{LAMA, Universit\'e Savoie Mont Blanc, France}}
\\[0.08\textheight]

\vspace*{1.1cm}

\colorbox{Lightblue}{
  \parbox[t]{1.0\textwidth}{
    \centering\huge\sc
    \vspace*{0.7cm}
    
    \textcolor{bluepigment}{On the modelling of shallow turbidity flows}
    
    \vspace*{0.7cm}
  }
}

\vfill % Whitespace between the title block and the publisher

\raggedleft     % Right-align all text
{\large \plogo} % Publisher and logo
\end{titlepage}

%%% ----------------------------------------------------------------------- %%%

\newpage
\thispagestyle{empty} % Remove page numbering on this page
\par\vspace*{\fill}   % Whitespace until the bottom
\begin{flushright} % Right-align all text
{\textcolor{denimblue}{\textsc{Last modified:}} \today}
\end{flushright}

%%% ----------------------------------------------------------------------- %%%

\newpage
\maketitle
\thispagestyle{empty}

%%% ----------------------------------------------------------------------- %%%

\begin{abstract}

In this study we investigate shallow turbidity density currents and underflows from mechanical point of view. We propose a simple hyperbolic model for such flows. On one hand, our model is based on very basic conservation principles. On the other hand, the turbulent nature of the flow is also taken into account through the energy dissipation mechanism. Moreover, the mixing with the pure water along with sediments entrainment and deposition processes are considered, which makes the problem dynamically interesting. One of the main advantages of our model is that it requires the specification of only two modeling parameters --- the rate of turbulent dissipation and the rate of the pure water entrainment. Consequently, the resulting model turns out to be very simple and self-consistent. This model is validated against several experimental data and several special classes of solutions (such as travelling, self-similar and steady) are constructed. Unsteady simulations show that some special solutions are realized as asymptotic long time states of dynamic trajectories.

\bigskip
\noindent \textbf{\keywordsname:} turbidity currents; density flows; shallow water flows; conservation laws; finite volumes; travelling waves; self-similar solutions \\


\end{abstract}

%%% ----------------------------------------------------------------------- %%%

\newpage
\tableofcontents
\thispagestyle{empty}

%%% ----------------------------------------------------------------------- %%%

\newpage
\section{Introduction}

Underwater turbidity currents are sediment-laden underflows that play an important r\^ole in the morphology of the continental shelves (more generally of ocean bottoms) and in the global sediment cycle going to the formation of hydrocarbon reservoirs. We refer to \cite{Ungarish2009} for a self-contained and comprehensive account of the theory of gravity currents and intrusions. The presence and entrainment of sediments differentiates them from stratified flows due to, \eg temperature or salinity differences. The main physical mechanisms include the deposition, erosion and dispersion of important amounts of heavy sediment particles. Turbidity currents are not to be confused with \emph{debris flows}, which represent fast-moving masses of poorly sorted heterogeneous material where interactions among the material pieces ($\approx$ particles) are important. Moreover, debris mix little with the ambient fluid. Debris flows have been a mainstream topic in the scientific literature due to their hazard they wreak in mountain regions (and not only).

The driving force is the gravity acceleration acting on dispersed sediment particles along steep and moderate bottom slopes. The initial perturbation is amplified by this acceleration, which in turn destabilizes the flow into shear instabilities that result in turbulent mixing and the transfer of mass and momentum. This gravity force creates the horizontal pressure gradient due to the increase of hydrostatic pressure resulting from the addition of particles. The heavy sediment particles are suspended in the mixing layer by fluid turbulence. The studied here processes are responsible of the transfer of littoral sediments to deep ocean regions. One should not disregard the destructive potential of gravity currents onto underwater structures such as pipelines, cables, \etc Turbidity currents in submarine canyons can attain surprisingly high velocities of the order of $8\ \sim\ 14\ \m/\s$ \cite{Krause1970, Parker1986}. These high velocities in the downstream direction result from the self-acceleration (and self-suspension) process from an appropriate initial perturbation, when more and more sediments are entrained by the flow from the bed, thus, increasing the rate of work performed by gravity \cite{Parker1986}. This process is sometimes referred to as the ``ignition'' \cite{Parker1982, Parker1986, Garcia1993}, which translates the energy imbalance property of such flows. One of important scientific questions is to determine the conditions necessarily to have an igniting flow. However, the self-acceleration stage cannot continue indefinitely. Most often the bed slope drops off (due to the bed morphology) or, simply, the sediment supply ceases. The mechanism of ignition was already described in \textsc{Pantin} (1979) \cite{Pantin1979}. However, the first laboratory demonstration of self-accelerated turbidity flows took $30$ more years \cite{Sequeiros2009}.

Turbidity currents is a particular case of (continuously) stratified flows and they are fundamentally different from classical density underflows \cite{Ellison1959}. The main difference comes from the fact that the source of the density gradient, \ie the suspended sediment, is not conservative. The suspended sediments are free to exchange with the core layer near the sea bed. The ambient still water is also entrained into this process. These exchanges are difficult to quantify and they constitute one of the main difficulties in the modeling of such flows \cite{Pantin1979, Parker1982, Parker1986}. In this respect turbidity currents are fundamentally non-conservative flows in their nature. Gravity flows may occur in the atmosphere\footnote{For instance, downslope windstorms over topography in \textsc{Colorado} (US) were observed and examined in \cite{Lilly1978, Neiman1988}.} over topography, sub-aerial (\eg avalanches, pyroclastic flows) and sub-aqueous environments (\eg turbidity currents) over bathymetries. They may result also from anthropogenic activities such as when a dense buoyant industrial effluent or pollutant is released into a lake, river or ocean. In the present study we shall consider mainly sub-aqueous flows due to the abundance of available experimental data. We refer to \cite{Parsons2007, Meiburg2010} as general excellent reviews on this topic.

Perhaps, the first serious attempts to observe turbidity currents in natural environments were performed in late 1960's at \textsc{Scripps} Canyon offshore of \textsc{La Jolla}, \textsc{California}. They were reported in \cite{Inman1976}. However, the flows reported in that study were so violent that the instrumentation was lost during these density currents making the detailed analysis extremely difficult \cite{Parsons2007}. The exact time moment of these underwater events is unpredictable which make them difficult to monitor in natural environments. Most of our physical knowledge on underwater turbidity currents come from small scale laboratory experiments \cite{Middleton1967, Garcia1993, Kubo2002, Kubo2004}. The experiments are bound to use common liquids for practical reasons. In general, it is not possible to respect all scalings. To give an example, we can mention the issue with particle sizes and their settling velocity. Nevertheless, taking into account the difficulties in obtaining field data, laboratory experiments are the only source of quantitative data about turbidity currents. The mathematical modelling is needed to extrapolate these experimental results to the scales on which these processes occur in nature. Nonetheless, the experiments offer a great opportunity for the verification of numerical results.

The gravity current can be divided geometrically into the flow \emph{head}, \emph{body} and \emph{tail}. The head is shaped as an ellipse and, generally the head is higher than the flow body. In the present study we are mainly interested in the flow head modelling, where the most intensive mixing processes take place. Consequently, it influences the whole flow dynamics. The most advanced point of the flow head is called the front or nose.

The main difficulties in understanding the dynamics of gravity turbidity currents come from their genuinely turbulent nature. Moreover, the phenomenon is nonlinear, heterogeneous and unsteady. The flow complexity increases when the flow entrains more and more sediments in suspension. The literature devoted to the mathematical modeling of the density currents is abundant. First of all, we would like to mention the classical monographs on this subject \cite{Turner1973, Townsend1980, Liapidevskii2000}. The first and simplest models intended to explain the classical lock-exchange configurations was proposed in \cite{Huppert1980}. These models are referred to as \emph{integral}, \emph{box} or 0D models, since all quantities are averaged in space. The modern approaches to the mathematical modeling of such flows were initiated in \cite{Pantin1979, Parker1982, Parker1986}. A dense cloud 0D model for powder-snow avalanches including non-\textsc{Boussinesq} and sediment entrainment effects along the avalanche path was proposed in \cite{Rastello2004}. Powder-snow avalanches are large-scale, finite volume release turbidity currents (in the form of large scale suspension clouds) occurring on mountain slopes. These clouds sometimes reach $100\ \m$ in height and the front velocities of the order of $100\ \frac{\m}{\s}\,$. Without sediments (\ie snow in the case of avalanches) distributed over the incline, the density current first accelerates and then decelerates without reaching important velocities. With sediments entrainment, the current can be maintained in the accelerating self-sustaining state during sufficient intervals of time to reach the velocities indicated above. In \cite{Hopfinger1983} a fair correlation of the avalanche velocity with the snow cover was demonstrated. The measurements of an avalanche front velocity in the \textsc{Sion} valley, \textsc{Switzerland} demonstrate a constant increase of the front velocity with traveled distance \cite{Dufour2001} (during the accelerating phase, of course). Thus, we come to the conclusion that the inclusion of sediments entrainment effect is of capital importance to predict the correct density current front velocity.

Some of recent studies devoted to the sediments transport within depth-averaged models include \cite{Bradford1999, Khan2005, Fernandez-Nieto2007, MoralesdeLuna2009, Benkhaldoun2009a}. This list is far from being exhaustive. The shallow water approach assumes that vertical accelerations are negligible, so the pressure being essentially hydrostatic. The sediment concentration is a passive tracer with exchanges among different layers. The flow is fully turbulent, even if pure viscous effects are generally negligible. Moreover, the energy required to keep the sediments in the suspension cloud is a negligible portion of the total turbulent energy production \cite{Parker1986}. Thus, the base model has to be first \textsc{Reynolds}-averaged \cite{Sreenivasan1999} before applying the long wave approximation. Several authors made an effort to take into account the turbulence modeling into the shallow water type models \cite{Mei2003, Fe2008, Fe2009}. Our approach to solve this issue will be detailed below. Nowadays, the multi-layer approaches to the density stratified flows become more and more popular \cite{Audusse2010}. Finally, some researchers chose a more CFD\footnote{Computational Fluid Dynamics (CFD).}-like approach to the simulation of density flows incorporating eventually the advanced turbulence modeling \cite{Birman2005, Etienne2005, Ozgokmen2006, Ozgokmen2007}. Perhaps, the first Direct Numerical Simulation (DNS) of the gravity current dates back to the years of $2\,000$ \cite{Hartel2000}. These simulations have an advantage of being depth-resolving and, thus, providing very a complete information about the flow structure in two or even three dimensions. However, due to the high computational complexity, only idealized academic configurations can be considered within reasonable CPU-time at the current state of technologies. Recently proposed three-dimensional (3D) turbidity-current models can be found in \cite{Imran2004, Kassem2004, Huang2005}. Moreover, the 3D DNS computations are often limited in the bulk \textsc{Reynolds} number.

In the present study we adopt a simplified (1.5D) approach along the lines of \cite{Salaheldin2000, Choi2001, Liapidevskii2004} based on the \textsc{Eulerian} formulation and depth-averaged formulations. A \textsc{Lagrangian} simplified \textsc{BANG1D} model was proposed in \cite{Pratson2001}. A simple 1.5D model was proposed in \cite{Johnson2013a}. The authors parametrized their model by making the entrainment velocity depending on the dimensionless \textsc{Richardson} number. In the present study we close the model in an alternative way.

Very similar physical processes take place in powder-snow avalanches where the snow particles suspension flows down the mountains and the snow plays the r\^ole of sediments in underflows \cite{Hopfinger1983}. Consequently, very similar mathematical models appear in these two fields and to make the bibliography review more complete we mention some recent results in powder-snow avalanche modeling \cite{Naaim1998, Blagovechshenskiy2002, Ancey2004, Etienne2005, Turnbull2007, Meyapin2009, Dutykh2009}.

The goal here is to propose a simple multi-layer (just two or three layers) shallow water-type model which takes into account mixing processes between the layers. We assume that the sediment particles are well mixed across the height of each layer. So that their volume fraction can be effectively approximated by depth-averaged quantities. This model preferably has to be simple enough to be studied using even analytical methods. The main scientific question which is currently poorly understood is the influence of flow stratification on global flow patterns. For instance, the formation of current's head (or the front region) and its steady-state velocity have to be carefully explained \cite{Liapidevskii2004}.

The present study is organized as follows. In Section~\ref{sec:model} we state the physical and mathematical problem under consideration. In Section~\ref{sec:analytical} we study the proposed model analytically and in Section~\ref{sec:exp} we validate its predictions by comparing them against experimental data. Some unsteady simulations and their relation to analytical (self-similar) solutions are presented in Section~\ref{sec:exp} as well. Finally, the main conclusions and perspectives of this study are outlined in Section~\ref{sec:disc}.

%%% ----------------------------------------------------------------------- %%%

\section{Mathematical model}
\label{sec:model}

Consider an incompressible liquid which fills a two-dimensional fluid domain $\Omega$. A Cartesian coordinate system $O\,\x$, $\x\ =\ (x,\,y)$ is chosen in a classical way such that the axis $Oy$ points vertically upward and abscissa $O\,x$ is positive along the right horizontal direction. The fluid is bounded below by a solid non-erodible bottom $y\ =\ d\,(x)\,$. Above, the fluid can be assumed unbounded for the sake of simplicity, since our attention will be focused on processes taking place in the region close to the bottom.

\begin{figure}
  \centering
  \includegraphics[width=1.0\textwidth]{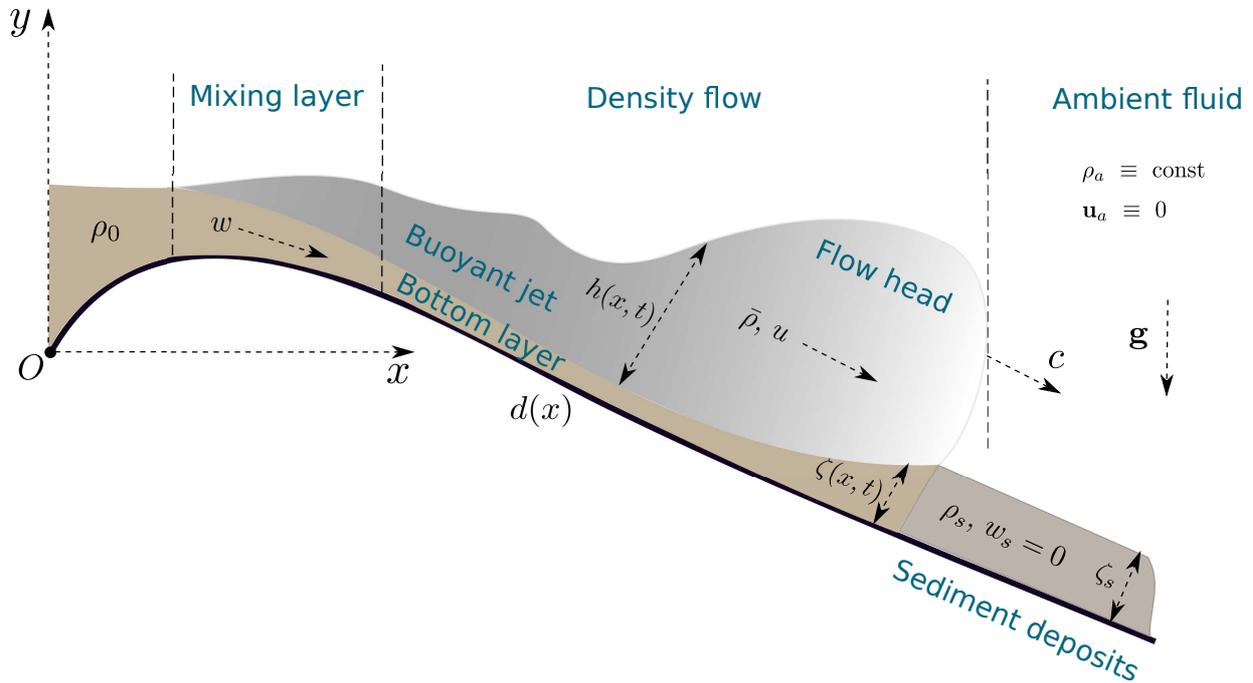}
  \caption{\small\em Sketch of the physical domain and the flow configuration. Various notations and variables are introduced and explained in the text.}
  \label{fig:sketch}
\end{figure}

The fluid is inhomogeneous and the flow can be conventionally divided in three parts. On the solid bottom there is a heavy fluid layer with density $\rho_{\,0}$ composed mainly of sedimentary deposits. Its thickness is $\zeta\,(x,\,t)\,$. Above, we have a muddle layer whose density will be denoted by $\rhob\,(\x,\,t)$ composed of the sediments and the still water mix. Its thickness is $h\,(x,\,t)\,$. Finally, the whole domain above these two layers is filled with the still ambient water of the density $\rho_{\,a}\ >\ 0$ at rest ($\u_{\,a}\ \equiv\ \vO$). This situation is schematically depicted in Figure~\ref{fig:sketch}. We include also into consideration the situation where a thin motionless layer of sediments with constant thickness $\zeta_{\,s}$ and density $\rho_{\,s}\ >\ 0$ covers the slope. It is usually the case in many practical situations and these sediment deposits may contribute to the flow head dynamics while propagating along the incline. One may imagine also that a certain mass of a sediment suspension or dense fluid is released into the flow domain at $x\ =\ 0$ with the mass flux $\rho_{\,0}\cdot\zeta\,(0,\,t)\cdot u\,(0,\,t)\,$.

\begin{table}
  \centering
  \begin{tabular}{c|l}
  \hline\hline
  \textit{Variable} & \textit{Significance} \\
  \hline\hline
  $b_{\,s}$ & Buoyancy of sediments \\
  $b_{\,0}$ & Buoyancy of the lower (bottom) layer \\
  $b\,(x,\,t)$ & Buoyancy in the mixing layer \\
  $\zeta_{\,s}$ & Thickness of the sediment deposit \\
  $\zeta\,(x,\,t)$ & Thickness of the lower (bottom) layer \\
  $h\,(x,\,t)$ & Thickness of the mixing layer \\
  $w\,(x,\,t)$ & Depth-averaged mean velocity in the bottom layer \\
  $u\,(x,\,t)$ & Depth-averaged mean velocity in the mixing layer \\
  $q\,(x,\,t)$ & Depth-averaged turbulent mean squared velocity in the mixing layer \\
  $\E\,(x,\,t)$ & Turbulent energy dissipation \\
  $u_{\,\ast}$ & Friction velocity \\
  \hline\hline
  \end{tabular}
  \bigskip
  \caption{\small\em Definitions of various parameters used in the approximate models. For the illustration see Figure~\ref{fig:sketch}. A more complete list of employed nomenclature is given in~\ref{app:nom}.}
  \label{tab:defs}
\end{table}

After applying the standard long wave scaling (or, equivalently, the depth-averaging), one can derive the following system of equations (\cf \cite{Liapidevskii2004}):
\begin{align}\label{eq:nc1}
  \zeta_{\,t}\ +\ [\,\zeta\,w\,]_{\,x}\ =&\ \chi^{\,-}\,, \\
  h_{\,t}\ +\ [\,h\,u\,]_{\,x}\ =&\ \chi^{\,+}\,, \label{eq:nc2} \\
  w_{\,t}\ +\ w\,w_{\,x}\ +\ [\,b_{\,0}\,\zeta\ +\ b\,h\,]_{\,x}\ =&\ -b_{\,0}\,d_{\,x}\ -\ \frac{u_{\,\ast}^{\,2}}{\zeta}\,, \label{eq:nc3} \\
  b_{\,t}\ +\ u\,b_{\,x}\ =&\ -\frac{b_{\,0}\,\chi^{\,-}\ +\ b\,\chi^{\,+}}{h}\,, \label{eq:nc4} \\
  u_{\,t}\ + u\,u_{\,x}\ +\ b\,[\,\zeta\ +\ h\,]_{\,x}\ +\ \half\,h\,b_{\,x}\ =&\
  -\ \frac{w\,\chi^{\,-}\ +\ u\,\chi^{\,+}}{h} - b\,d_{\,x}\,, \label{eq:nc5} \\
  q_{\,t}\ +\ u\,q_{\,x}\ =&\ (2\,h\,q)^{\,-1}\Bigl[\,\bigl(2\,w\,u\ -\ w^{\,2}\ +\ b_{\,0}\,h\ -\ 2\,b\,h\bigr)\,\chi^{\,-} \nonumber \\
  & +\ \bigl(u^{\,2}\ -\ q^{\,2}\ -\ b\,h\bigr)\,\chi^{\,+}\ -\ \E\,\Bigr]\,.\label{eq:nc6}
\end{align}
The sense of the variables is explained in Table~\ref{tab:defs} and also in Figure~\ref{fig:sketch}. For the dissipative term, we assume the following closure relation:
\begin{equation*}
  \E\ \eqdef\ \kappa\,q^{\,3}\,, \qquad \mbox{ where } \qquad \kappa\ >\ 0\,.
\end{equation*}

In the sequel we are going to make the following choice of the entrainment rates $\chi^{\,\pm}$ (unless the contrary is stated explicitly):
\begin{equation*}
  \chi^-\ \equiv\ 0\,, \qquad \chi^+\ \equiv\ \sigma\,q\,.
\end{equation*}
Moreover, we assume that the gravity force balances exactly the turbulent friction at the solid bottom, \ie
\begin{equation*}%\label{eq:frict}
  \frac{u_{\,\ast}^{\,2}}{\zeta}\ \equiv\ -b_{\,0}\,d_{\,x}\,,
\end{equation*}
which implies that the kinematic Equation \eqref{eq:nc1} is completely decoupled from the rest of the system \eqref{eq:nc2} -- \eqref{eq:nc6}. The information about $\zeta$ is transported along characteristics of this equation. So, if $\zeta$ is constant initially and this constant value is maintained at the channel inflow, $\zeta$ will remain so under the system dynamics. We will adopt this assumption as well in order to focus our attention on the mixing layer dynamics. Below we will consider only the middle mixing layer. Under these conditions, the equilibrium model becomes:
\begin{align}\label{eq:eq1}
  h_{\,t}\ +\ [\,h\,u\,]_x\ =&\ \sigma\,q\,, \\
  b_{\,t}\ +\ u\, b_{\,x}\ =&\ -\frac{\sigma\, q\, b}{h}\,, \\
  u_{\,t}\ +\ u\,u_{\,x}\ +\ b\, h_{\,x}\ +\ \half\, h\, b_{\,x}\ =&\ -\frac{\sigma\, q\, u}{h}\ -\ b\, d_{\,x}\,, \\
  q_{\,t}\ +\ u\,q_{\,x}\ =&\ \sigma\;\frac{u^{\,2}\ -\ q^{\,2}\ - h\, b\ -\ \delta q^{\,2}}{2\,h}\,,\label{eq:eq4}
\end{align}
where the constant $\delta$ is defined as: 
\begin{equation*}
  \delta\ \eqdef\ \frac{\kappa}{\sigma}\ >\ 0\,.
\end{equation*}
The system \eqref{eq:eq1} -- \eqref{eq:eq4} can be equivalently recast in the conservative form which has an advantage to be valid for discontinuous solutions as well \cite{Rozhdestvenskiy1978, Godlewski1990}:
\begin{align}\label{eq:cons1}
  h_{\,t}\ +\ [\,h\,u\,]_{\,x}\ =&\ \sigma\,q\,, \\
  (h\,b)_{\,t}\ +\ [\,h\,b\,u\,]_{\,x}\ =&\ 0\,, \\
  (h\,u)_{\,t}\ +\ \bigl[\,h\,u^{\,2}\ +\ \half\, b\, h^{\,2}\,\bigr]_{\,x}\ =&\ -h\,b\,d_{\,x}\,, \\
  \bigl(h\,(u^{\,2}\ +\ q^{\,2}\ + h\,b)\bigr)_{\,t}\ +\ \bigl[\,(u^{\,2}\ +\ q^{\,2}\ +\ 2\,h\,b)\;h\,u\,\bigr]_{\,x}\ =&\ -2\,h\,b\,u\,d_{\,x}\ -\ \kappa\,q^{\,3}\,.\label{eq:cons4}
\end{align}
Here, $\kappa\ \in\ \R^{\,+}$ is a positive constant measuring the rate of turbulent dissipation \cite{Sreenivasan1999, Lesieur2008}.

An important property of this model is the absence of the bottom friction term. This physical effect was taken into account, while excluding the bottom layer of thickness $\zeta(x,\,t)\,$. This gives us the mathematical reason for the absence of a friction term in model \eqref{eq:cons1} -- \eqref{eq:cons4}. Similarly, we can bring also a physical argument to support this fact. The horizontal velocity takes the maximum value on the boundary between the bottom sediment and mixing layers. Consequently, the \textsc{Reynolds} stress $\tau_\ast$ vanishes here.

\begin{remark}
In fact, we can derive an additional balance law by combining together Equations \eqref{eq:eq1} and \eqref{eq:eq4}:
\begin{equation*}
  (h\,q)_t\ +\ [\,h\,q\,u\,]_x\ =\ \half\,\sigma\;\bigl(u^2\ +\ (1-\delta)\,q^2\ -\ h\,b\bigr)\,.
\end{equation*}
The last equation is not independent from Equations \eqref{eq:cons1} -- \eqref{eq:cons4}. Consequently, it does not bring new information about the equilibrium sedimentation-free model \eqref{eq:eq1} -- \eqref{eq:eq4}. Nevertheless, we provide it here for the sake of the exposition completeness.
\end{remark}

The proposed model has an advantage of being simple, almost physically self-consistent\footnote{Only two constants $\sigma$ and $\delta$ need to be prescribed to close the system.} and having the hyperbolic structure. It was derived for the first time in \cite[Chapter~5]{Liapidevskii2000}. In order to obtain a well-posed problem the system \eqref{eq:cons1} -- \eqref{eq:cons4} has to be completed by corresponding boundary and initial conditions. In the following sections, the just proposed equilibrium system \eqref{eq:cons1} -- \eqref{eq:cons4} will be studied in more details by analytical and numerical means.

%%% ----------------------------------------------------------------------- %%%

\section{Analytical study of the model}
\label{sec:analytical}

In this Section we discuss the steady state solutions and investigate the qualitative behaviour of two important special classes of solutions --- travelling waves and similarity solutions.

%%% ----------------------------------------------------------------------- %%%

\subsection{Mixing layer formation}
\label{sec:mixing}

Mixing layers are formed between two fluid layers having different densities. The initial stages of mixing processes play a very important r\^ole in natural and laboratory environments. The physical mechanism of a mixing layer formation is given by various interfacial instabilities (\eg \textsc{Rayleigh}--\textsc{Taylor} or \textsc{Kelvin}--\textsc{Helmholtz}). The most important mechanism is, of course, the \textsc{Kelvin}--\textsc{Helmholtz} instability \cite{Helmholtz1868} because of the shear velocity presence \cite{Pawlak1998}. In supercritical flows the mixing intensity is considerably intensified. In this way, mixing layers are the most pronounced in transcritical flows over underwater obstacles (\ie bathymetric features), since a jet is formed on the obstacle downstream (lee) side making the flow supercritical. The process of mixing layer formation over an inclined bottom was studied in \cite{Liapidevskii2004} in the framework of a three-layer model. In the present study we apply a similar approach to the two-layer\footnote{We remind that the upper layer of the still water is assumed to be motionless in the present study.} System \eqref{eq:nc1} -- \eqref{eq:nc6}. Below we derive a steady solution, which provides the boundary conditions for the unsteady propagation of the flow head (see Figure~\ref{fig:sketch}).

Stationary solutions to System \eqref{eq:nc1} -- \eqref{eq:nc6} satisfy the following system of differential equations:
\begin{align*}
  w\,\zeta_{\,x}\ +\ \zeta\,w_{\,x}\ &=\ \chi^{\,-}\,, \\
  u\,h_{\,x}\ +\ h\,u_{\,x}\ &=\ \chi^{\,+}\,, \\
  w\,w_{\,x}\ +\ b_{\,0}\,\zeta_{\,x}\ +\ b\,h_{\,x}\ +\ h\,b_{\,x}\ &=\ -b_{\,0}\,d_{\,x}\ -\ \frac{\tau_{\,\ast}}{\zeta}\,, \\
  u\,u_{\,x}\ +\ b\,(\zeta_{\,x}\ +\ h_{\,x})\ +\ \half\,h\,b_{\,x}\ &=\ -\,b\,d_{\,x}\ -\ \frac{w\,\chi^{\,-}\ +\ u\,\chi^{\,+}}{h}\,, \\
  h\,u\,b_{\,x}\ &=\ -b_{\,0}\,\chi^{\,-}\ -\ b\,\chi^{\,+}\,, \\
  2\,q\,h\,u\,q_{\,x}\ &=\ \bigl(2\,w\,u\ -\ w^{\,2}\ +\ b_{\,0}\,h\ -\ 2\,b\,h\bigr)\,\chi^{\,-}\\
  & +\ \bigl(u^{\,2}\ -\ q^{\,2}\ -\ b\,h\bigr)\,\chi^{\,+}\ -\ \kappa\,q^{\,3}\,.
\end{align*}
The last system of equations can be seen as a quasilinear system with respect to spatial derivatives of unknown variables $\bigl(h_{\,x},\,u_{\,x},\,b_{\,x},\,\zeta_{\,x},\,w_{\,x},\,q_{\,x}\bigr)\,$. The determinant $\Delta\,(x)$ of this system can be easily computed:
\begin{equation}\label{eq:det}
  \Delta\,(x)\ =\ \bigl(u^{\,2}\ -\ b\,h\bigr)\cdot\bigl(w^{\,2}\ -\ b_{\,0}\,\zeta\bigr)\ -\ b^{\,2}\,\zeta\,h\,.
\end{equation}
By assuming that $\Delta\,(x)\ \neq\ 0\,$, we obtain:
\begin{align}\label{eq:steady1}
  h_{\,x}\ &=\ \frac{a_{\,1}\cdot b\,h\ +\ a_{\,2}\cdot\bigl(w^{\,2}\ -\ b_{\,0}\,\zeta\bigr)}{\Delta\,(x)}\,, \\
  u_{\,x}\ &=\ \frac{\chi^{\,+}\ -\ u\,h_{\,x}}{h}\,, \quad (h_x \mbox{ should be taken from the previous equation}) \label{eq:steady2} \\
  b_{\,x}\ &=\ -\,\frac{b_{\,0}\,\chi^{\,-}\ +\ b\,\chi^{\,+}}{h\,u}\,, \label{eq:steady3} \\
  \zeta_{\,x}\ &=\ -\,\frac{1}{b}\;\Bigl(\half\,h\,b_{\,x}\ +\ \frac{w\,\chi^{\,-}\ +\ u\,\chi^{\,+}}{h}\ +\ u\,u_{\,x}\Bigr)\ -\ h_{\,x}\ -\ d_{\,x}\,, \label{eq:steady4} \\
  w_{\,x}\ &=\ \frac{\chi^{\,-}\ -\ w\,\zeta_{\,x}}{\zeta}\,, \label{eq:steady5} \\
  q_{\,x}\ &=\ \frac{\bigl(2\,w\,u\ -\ w^{\,2}\ +\ b_{\,0}\,h\ -\ 2\,b\,h\bigr)\,\chi^{\,-}\ +\ \bigl(u^{\,2}\ -\ q^{\,2}\ -\ b\,h\bigr)\,\chi^{\,+}\ -\ \kappa\,q^{\,3}}{2\,q\,h\,u}\,, \label{eq:steady6}
\end{align}
where the coefficients $a_{\,1,\,2}$ are defined as
\begin{align*}
  a_{\,1}\ &\eqdef\ \chi^{\,-}\,w\ +\ b_{\,0}\,\zeta\,d_{\,x}\ +\ \tau_{\ast}\,, \\
  a_{\,2}\ &\eqdef\ \chi^{\,+}\,u\ +\ b\,h\,d_{\,x}\ +\ w\,\chi^{\,-}\ +\ u\,\chi^{\,+}\,.
\end{align*}
The last system of equations can be seen as an Initial Value Problem (IVP) in the spatial variable $x\,$, if all spatial derivatives in the right-hand side are replaced by their expressions (we do not make this operation to keep the shorthand notation). We illustrate below the behaviour of solutions to System \eqref{eq:steady1} -- \eqref{eq:steady6} on the example of the mixing layer formation problem over an inclined bottom. The real data to build this solution were taken from \cite{Pawlak2000}.

%%% ----------------------------------------------------------------------- %%%

\subsubsection{Problem statement and solution}

In this Section we formulate the IVP inspired by the experimental study \cite{Pawlak2000}. Consider a density current over the flat plane inclined with angle $\phi$ with respect to the horizontal direction. Without any loss of generality, we postulate that the mixing layer starts to form at some location $x\ =\ x_{\,0}\,$, where we set the initial conditions for the steady System \eqref{eq:steady1} -- \eqref{eq:steady6}. Moreover, we adopt the following closure: 
\begin{equation}\label{eq:ml}
  \chi^{\,+}\ =\ 2\,\sigma\,q\,, \qquad
  \chi^{\,-}\ =\ -\,\sigma\,q\,.
\end{equation}
We assume that in this point we have a supercritical flow with velocity $w_{\,0}\,$, height $\zeta_{\,0}$ and density $\rho_{\,0}\,$. The upper layer of still water with density $\rho_{\,a}$ is motionless. Thus, we have
\begin{equation*}
  w_{\,0}^{\,2}\ >\ b_{\,0}\,\zeta_{\,0}\,,
\end{equation*}
and $b_{\,0}$ is defined as 
\begin{equation*}
  b_{\,0}\ \eqdef\ \frac{\bigl(\rho_{\,0}\ -\ \rho_{\,a}\bigr)\,g}{\rho_{\,a}}\,.
\end{equation*}
The initial height of the mixing layer is $h_{\,0}\ =\ 0$ by our assumption on point $x_{\,0}\,$. The initial asymptotic stages of the mixing layer formation as $x\ \to\ x_{\,0}$ can be determined from the condition that the right hand sides in \eqref{eq:steady1} -- \eqref{eq:steady6} are bounded \cite{Liapidevskii2000, Liapidevskii2004}:
\begin{equation*}
  b_{\,L}\ =\ \half\, b_{\,0}\,, \qquad
  u_{\,L}\ =\ \half\, w_{\,0}\,, \qquad
  q_{\,L}\ =\ \frac{w_{\,0}}{\sqrt{4\ +\ 2\,\delta}}\,.
\end{equation*}
Under the condition $\zeta\ >\ 0\,$, we shall have $b\,(x)\ \equiv\ b_{\,L}\ =\ \half\, b_{\,0}\,$. This can be seen after substituting the closure relations for $\chi^{\,\pm}$ from Equation~\eqref{eq:ml} into Equation~\eqref{eq:steady3}. One readily obtains that
\begin{equation*}
  b_{\,x}\ =\ \frac{\sigma\,q\,\bigl(b_{\,0}\ -\ 2\,b\bigr)}{h\,u}\,.
\end{equation*}
From the condition $b_{\,L}\ =\ \half\, b_{\,0}\,$, it follows that $b\,(x)\ \equiv\ \half\, b_{\,0}\,$.

\begin{figure}
  \centering
  \includegraphics[width=0.99\textwidth]{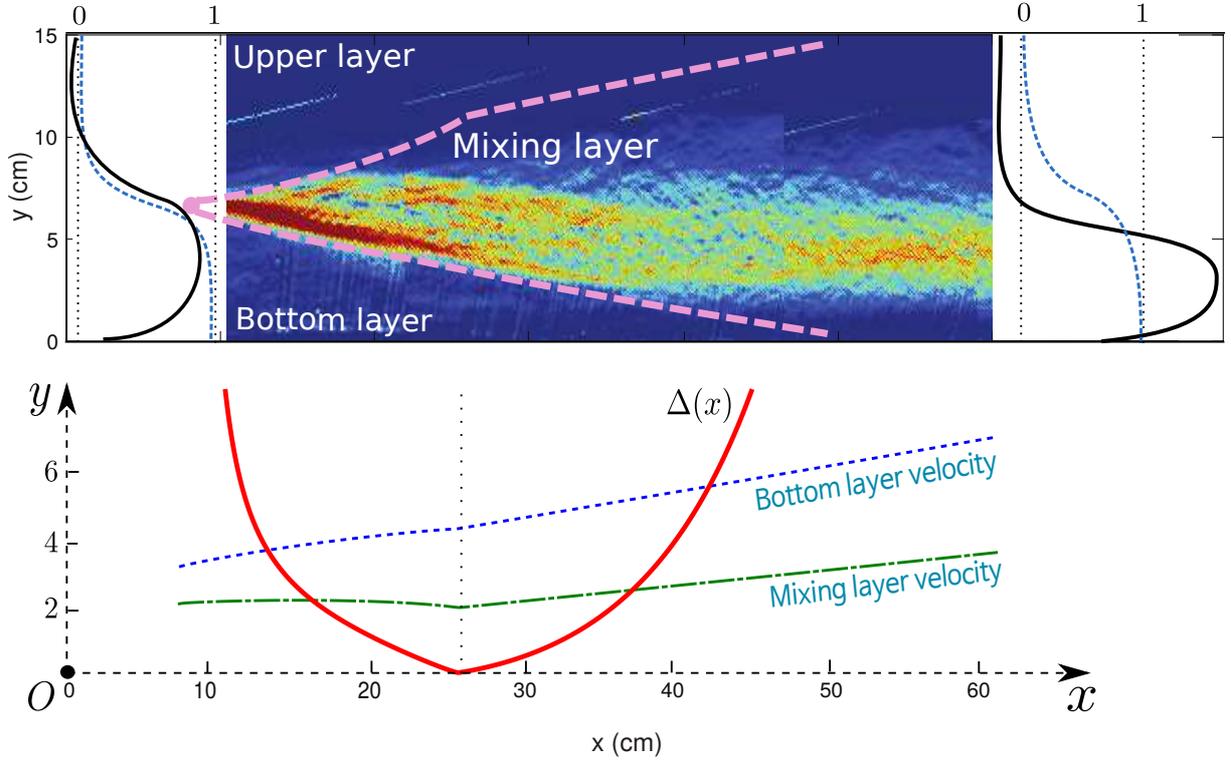}
  \caption{\small\em The density gradient field image: comparison with the experimental measurements using Laser Induced Fluorescence (LIF) \cite{Pawlak2000}. The picture is rotated so that the slope becomes horizontal.}
  \label{fig:pawlak}
\end{figure}

By taking into account the last asymptotics for the solution, we can construct it using standard numerical means until the point $x_{\,1}\,$, where the mixing layer will touch the bottom, \ie the point where $\zeta(x_{\,1})\ \equiv\ 0\,$. The solution to system \eqref{eq:steady1} -- \eqref{eq:steady6} taking into account the aforementioned asymptotics is depicted in Figure~\ref{fig:pawlak}. We used the following set of physical parameters in our computations:
\begin{multline*}
  \sigma\ =\ 0.15\,, \qquad
  \delta\ =\ 6\,, \qquad
  \phi\ =\ 10.8^{\,\circ}\,, \qquad
  x_{\,0}\ =\ 8\,\mathsf{cm}\,, \\
  \zeta_{\,0}\ =\ 7\,\mathsf{cm}\,, \qquad
  w_{\,0}\ =\ 5\;\frac{\mathsf{cm}}{\mathsf{s}}\,, \qquad
  b_{\,0}\ =\ 1.4\;\mathsf{\frac{cm}{s^{\,2}}}\,.
\end{multline*}
The flow geometry along with the initial conditions were taken from \cite{Pawlak2000}. On the upper panel of Figure~\ref{fig:pawlak} we show the comparison with experimental data provided by \textsc{Armi} \& \textsc{Pawlak} (2000) \cite{Pawlak2000}, who employed the Laser Induced Fluorescence (LIF) and Digital Particle Imaging Velocimetry (DPIV) together with a concentration of Rhodamine dye for flow visualization. Color code corresponds to the density gradient (blue is lower, red is higher). Their results show that instabilities evolve in an asymmetrical manner. In Figure~\ref{fig:pawlak} with a dashed line we show the numerical prediction given by function $y\ =\ \zeta\,(x)$ (lower line) and $y\ =\ \zeta\,(x)\ +\ h\,(x)$ (upper line). The reasonable agreement between the experimental data and our numerical solution at the initial parts of the mixing layer validate the approximate model (at least for steady solutions). We would like to mention that in the experiment (as well as in the nature), there is a slight backward flow above the mixing layer. The typical velocities are of the order $0.5\ \sim\ 1\;\mathsf{\frac{\cm}{\s}}\,$. In our model this effect is not taken into account. It could be done by including the third (upper) layer into consideration \cite{Liapidevskii2004}. On the left and right sub-plots in upper Figure~\ref{fig:pawlak}, we show the experimental distribution of the velocity (solid line) and of the density (dashed line) at the beginning and at the end of the considered fluid domain.

The theoretical lower bound of the mixing layer coincides fairly well with the experimental estimation of the high density gradient area as it can be seen in Figure~\ref{fig:pawlak} (upper panel). On the other hand, the theoretical upper bound goes sufficiently higher than the coloured region measured experimentally. This little discrepancy comes from the fact that the experimental data report the high density gradient, while our model predicts the upper bound of large vortices appearing during nonlinear stages of the \textsc{Kelvin}--\textsc{Helmholtz} (KH) instability development \cite{Helmholtz1868}. This situation is described in more details in \cite[Chapter~7, \textsection~3]{Liapidevskii2000}. The experimental evidence for the KH mechanism is shown in \cite[Figure~8]{Pawlak2000}. In particular, it is shown there that the mixing layer growth into the still water is
\begin{equation*}
  h_{\,x}\ \equiv\ \od{h}{x}\ =\ 2\,\sigma\ \equiv\ 0.3 \quad (\mbox{in our computation}).
\end{equation*}
In the same time, the effective thickness of the mixing layer  grows according the computed velocity profile as
\begin{equation*}
  \od{h_{\,\mathrm{eff}}}{x}\ \approx\ 0.18\,.
\end{equation*}
This value corresponds much better to numerous experimental findings \cite[Fig.~7.4]{Liapidevskii2000}. Consequently, we may conclude that in the considered case of the stratified fluid flow down the incline, the \emph{effective} height\footnote{We remind here the definition of the effective height of the mixing layer: it is the fluid body bounded from above and below by virtual surfaces where $w\ =\ 0.95\,\times\,w_{\,0}$ and $w\ =\ 0.1\,\times\,w_{\,0}$ correspondingly. The asymmetry in this definition comes from the nonlinear form of the vertical velocity profile.} of the mixing layer will be smaller than the height of the computed turbulent layer.

In Figure~\ref{fig:pawlak} (lower panel) we show with dotted lines the (theoretically predicted) distribution of velocities $u\,(x)$ and $w\,(x)$ in the mixing layer. The slope rupture in these curves is visible. It happens since at $x_{\,c}\ \approx\ 26\;\textsf{cm}$ the determinant \eqref{eq:det} vanishes, \ie
\begin{equation*}
  \Delta\,(x_{\,c})\ \equiv\ 0\,.
\end{equation*}
Physically speaking it means that the flow remains everywhere supercritical. However, for small changes of the flow parameters, it is possible that hydraulic jumps will appear along with following local subcritical zones. We note also that internal hydraulic jumps on the downstream side of an underwater obstacle is an important feature of stratified flows (both in the ocean and in the atmosphere). The inclusion of the mixing layer formation before the hydraulic jump is of capital importance in such situations.

%%% ----------------------------------------------------------------------- %%%

\subsubsection{Sediment layer}

When the mixing layer approaches the lower boundary with sediment deposits, the mass entrainment from the lowest layer decelerates and stops completely. In the framework of our model, this effect could be realized by ensuring the transition from $\chi^{\,-}\ =\ -\,\sigma\,q$ towards $\chi^{\,-}\ =\ 0$ (and, correspondingly, from $\chi^{\,+}\ =\ 2\,\sigma\,q$ towards $\chi^{\,+}\ =\ \sigma\,q$). Physically it means that the maximum of the flow velocity is achieved somewhere near the boundary between the mixing layer and bottom layer of constant density $\rho_{\,0}\,$. The transition to $\chi^{\,-}\ =\ 0$ takes place in neighbourhood of the right boundary of the flow represented in Figure~\ref{fig:pawlak}.

%%% ----------------------------------------------------------------------- %%%

\subsubsection{Intermediate conclusions}

The mixing layer structure over a downhill determines the flow structure further down. In particular, the total mass flux (relative to fluid density $\rho_{\,a}$) $\M_{\,0}\ =\ b_{\,0}\,\zeta_{\,0}\,w_{\,0}$ is divided into two parts. Namely, this splitting takes place at the point of transition of the mixing layer into the turbulent jet and the undiluted bottom layer, \ie at $x\ =\ x_{\,1}\,$.

The turbulent jet receives the buoyancy flux $\M_{\,j}\ =\ b_{\,1}\,h_{\,1}\,u_{\,1}\,$, while the bottom layer takes $\M_{\,b}\ =\ b_{\,0}\,\zeta_{\,1}\,w_{\,1}\,$. We notice also that for the slope angle $\phi\ =\ 10.8^{\,\circ}$ and experimental facilities considered in Figure~\ref{fig:pawlak}, the dominant part of the mass flow goes into the turbulent jet, \ie $\M_{\,j}\ \gg\ \M_{\,b}\,$.

If we assume that the flow velocity at $x\ =\ x_{\,1}$ in the bottom layer stabilizes\footnote{Physically it happens when the gravity force projection along the slope is balanced by the friction force with the rigid bottom.} due to small thickness of the layer, in the bottom layer we have $w\ \equiv\ w_{\,b}\,$, $\zeta\ \equiv\ \zeta_{\,b}$ and the density in the layer does not change anymore. Hence, the bottom layer becomes completely passive in the flow dynamics. But in the same time it influences the flow head propagation velocity for moderate bottom slopes\footnote{For large inclines the physical mechanisms are slightly different.}. This point will be demonstrated below.

%%% ----------------------------------------------------------------------- %%%

\subsection{Steady flows}
\label{sec:steady}

From numerical and experimental points of view, a significant number of studies has been devoted to the study of the transient gravity current problem with special focus on the density front formation and evolution \cite{Britter1980, Hopfinger1983, Maxworthy2010}. The steady flow configuration has received less attention \cite{Garcia1993a}. However, when one has a dynamical system in hands, it is natural to begin its study by looking for equilibrium points \cite{Jordan2007}. Thus, the study of a mathematical model is not complete if we do not discuss this class of solutions. Moreover, stationary solutions may be \emph{realized} and \emph{observed} on certain time scales in laboratory experiments, if certain conditions are met and maintained during sufficiently long time. For instance, the buoyant flow entering the inclined channel has to be maintained at constant rate during the whole experiment. If permanent boundary conditions are maintained not only in the mixing layer, but also in the bottom layer as well, then, the lower sediment layer will play a passive r\^ole in the flow, \ie
\begin{equation*}
  \zeta_{\,b}\ =\ \const\,, \qquad w_{\,b}\ =\ \const\,, \qquad \rho_{\,0}\ =\ \const\,.
\end{equation*}
In other words, the friction and mixing between these two layers do not take place. It is not difficult to see that fluid flows of the form
\begin{equation*}
  u\ \equiv\ u_{\,j}\,, \qquad m\ =\ b\,h\ \equiv\ m_{\,j}\,, \qquad q\ \equiv\ q_{\,j}
\end{equation*}
satisfy Equations \eqref{eq:eq1} -- \eqref{eq:eq4} provided that the mixing layer thickness depends linearly on the coordinate $x$ along the channel, \ie
\begin{equation*}
  h\,(x)\ =\ h_{\,0}\ +\ \varsigma\cdot(x\ -\ x_{\,0})\,.
\end{equation*}
Here, quantities with the sub-script ${}_j$ denote values in the steady jet. The variable $m$ will be sometimes referred as the mass, however, physically it represents the excess of the fluid column weight with respect to the still water level due to the presence of heavy sediment suspensions. To satisfy the system \eqref{eq:eq1} -- \eqref{eq:eq4}, the following identities have to be satisfied:
\begin{align}\label{eq:1a}
  \varsigma\cdot u_{\,j}\ &=\ \sigma\cdot q_{\,j}\,, \\
  \varsigma\cdot \bigl(u_{\,j}^{\,2}\ +\ \half\,m_{\,j}\bigr)\ &=\ \alpha\cdot m_{\,j}\,,\label{eq:1b} \\
  \varsigma\cdot \bigl(u_{\,j}^{\,2}\ +\ q_{\,j}^{\,2}\ +\ 2\,m_{\,j}\bigr)\cdot u_{\,j}\ &=\ 2\,\alpha\cdot m_{\,j}\cdot u_{\,j}\ -\ \kappa\,q_{\,j}^{\,3}\,, \label{eq:1c}
\end{align}
where $\alpha\ \eqdef\ -\,d_{\,x}\ \geq\ 0\,$. An algebraic consequence of relations \eqref{eq:1a} -- \eqref{eq:1c} above can be easily derived:
\begin{equation*}
  u_{\,j}^{\,2}\ -\ m_{\,j}\ -\ (1\ +\ \delta)\,q_{\,j}^{\,2}\ =\ 0\,.
\end{equation*}
The boundary conditions specify also the buoyancy influx $\M_{\,j}$ into the channel:
\begin{equation*}
  \M_{\,j}\ \eqdef\ m_{\,j}\cdot u_{\,j}\ =\ U_{\,j}^{\,3}\ \quad \Longrightarrow\ \quad U_{\,j}\ =\ \sqrt[3]{\M_{\,j}}\,.
\end{equation*}
All the relations above can be combined into a single equation for the quantity $\a_{\,j}\ \eqdef\ \dfrac{u_{\,j}}{U_{\,j}}$ using simple algebraic transformations (see also~\ref{app:A1}):
\begin{equation*}
  \sqrt{\frac{1\ -\ \a_{\,j}^{\,-3}}{1\ +\ \delta}}\;\bigl(1\ +\ 2\,\a_{\,j}^{\,3}\bigr)\ =\ \frac{2\,\alpha}{\sigma}\,, \qquad \sigma\ \neq\ 0\,.
\end{equation*}
There exists a unique solution to this equation for any positive right hand side. Moreover, one can show that $\a_{\,j}\ >\ 1\,$. It corresponds to the supercritical flow with $u_{\,j}^{\,2}\ >\ m_{\,j}\,$. Then, once we determined $\a_{\,j}\,$, from Equations \eqref{eq:1a} -- \eqref{eq:1c} we can determine the remaining quantities $u_{\,j}\,$, $m_{\,j}\,$, $q_{\,j}$ and $\varsigma\,$.

To make a conclusion, in turbidity gravity flows in an inclined channel, where the horizontal (depth-integrated) velocity maximum is achieved on the boundary between the sediment and turbulent mixing layers, the model \eqref{eq:eq1} -- \eqref{eq:eq4} predicts the stationary \emph{supercritical} flow with pure fluid entrainment from the upper layer.

%%% ----------------------------------------------------------------------- %%%

\subsection{Travelling waves}
\label{sec:travel}

One of the main questions in the modelling of turbidity flows is to determine the density front velocity. The proposed base model \eqref{eq:eq1} -- \eqref{eq:eq4} is sufficiently simple to address this question analytically. 

%%% ----------------------------------------------------------------------- %%%

\subsubsection{General considerations}
\label{sec:gen}

In this Section we describe the class of travelling wave solutions to System \eqref{eq:cons1} -- \eqref{eq:cons4} in their generality. Since we are mainly interested in smooth solutions, we will use the non-conservative form \eqref{eq:eq1} -- \eqref{eq:eq4} for the sake of convenience. Also, we assume that the bottom slope $\alpha$ is constant, which is necessary for the existence of solutions with permanent shape.

The travelling wave ansatz takes the following form:
\begin{equation*}
  h\,(x,\,t)\ =\ h\,(\xi)\,, \quad m\,(x,\,t)\ =\ m\,(\xi)\,, \quad u\,(x,\,t)\ =\ u\,(\xi)\,, \quad
  q\,(x,\,t)\ =\ q\,(\xi)\,,
\end{equation*}
where $\xi\ \eqdef\ x\ -\ c\cdot t$ and $c$ is a positive\footnote{If the wave travels to the rightwards direction, as we assume without any loss of generality.} constant, which has the physical sense of the travelling wave speed (to be determined later). After substituting this ansatz into Equations \eqref{eq:eq1} -- \eqref{eq:eq4}, we obtain a system of Ordinary Differential Equations (ODEs):
\begin{eqnarray}
  (u\ -\ c)\,h^{\,\prime}\ +\ hu^{\,\prime}\ &=&\ \sigma\, q\,, \label{eq:ode1} \\
  (u\ -\ c)\,m^{\,\prime}\ +\ m\,u^{\,\prime}\ &=&\ 0\,, \label{eq:ode2} \\
  (u\ -\ c)\,u^{\,\prime}\ +\ \frac{1}{2}\,m^{\,\prime}\ +\ \frac{m}{2h}\;h^{\,\prime}\ &=&\ \frac{\alpha\,m\ -\ \sigma\, q\, u}{h}\,, \label{eq:ode3} \\
  (u\ -\ c)\,q^{\,\prime}\ &=&\ \frac{\sigma\bigl(u^{\,2}\ -\ (1\ +\ \delta)q^{\,2}\ -\ m\bigr)}{2\,h}\,, \label{eq:ode4}
\end{eqnarray}
where the prime ${}^{\prime}$ denotes the differentiation operation with respect to $\xi\,$. This system of ODEs describes a transcritical gravity flow in a coordinate system which moves with velocity $c\,$. More precisely, the flow is of supercritical type in the avalanche core and it switches to the subcritical regime when we cross the wave front. By analogy with the detonation theory \cite{Fickett1979}, the front velocity $c$ (under certain conditions) is given by the \textsc{Chapman}--\textsc{Jouguet} principle which states that a transcritical front propagates with the minimal admissible velocity \cite{Chapman1899}.

After some computations the ODEs System \eqref{eq:ode1} -- \eqref{eq:ode4} can be reduced to a single differential equation with implicit dependence on $\xi\,$:
\begin{equation}\label{eq:dyn}
  \od{q}{u}\ =\ \frac{\sigma}{2}\;\frac{\overbrace{\bigl(u^2\ -\ (1\ +\ \delta)\,q^2\ -\ m\bigr)}^{\displaystyle{\ \defeq\ \Aa\,(q,\,u)}}\cdot\overbrace{\bigl((u\ -\ c)^2\ -\ m\bigr)}^{\displaystyle{\ \defeq\ \Dd\,(u)}}}{(u\ -\ c)^2\cdot\underbrace{\Bigl(\alpha\,m\ -\ \sigma\, q\cdot \bigl(u\ +\ \frac{m}{2\,(u\ -\ c)}\bigr)\Bigr)}_{\displaystyle{\ \defeq\ \Bb\,(q,\,u)}}}\,,
\end{equation}
where the expression for $m\,(\xi)$ in terms of $u\,(\xi)$ is obtained by integrating Equation \eqref{eq:ode2}:
\begin{equation}\label{eq:m}
  m\,(\xi)\ =\ -\,\frac{\M\,(c)}{u\,(\xi)\ -\ c}\,.
\end{equation}
The integration ``constant'' $\M\,(c)$ can be determined from the \textsc{Rankine}--\textsc{Hugoniot} conditions written at the wave front \cite{Rozhdestvenskiy1978, Godlewski1990}.

%%% ----------------------------------------------------------------------- %%%

\subsubsection{Stability of equilibria}
\label{sec:stab}

In this Section we study the existence and stability of equilibria points to Equation~\ref{eq:dyn}. One of the difficulties is that the right hand side depends on a free parameter $\delta\,$, whose variation has to be taken into account. The wave celerity $c\ >\ u$ has to be specified as well.

As the first step, we rewrite Equation \eqref{eq:dyn} as a dynamical system in the plane $\bigl(q,\,u\bigr)\,$ with $\xi$ being the evolution variable:
\begin{align*}
  q^{\,\prime}\ &=\ \sigma\,\Aa\,(q,\,u)\,\Dd\,(u)\ \defeq\ \F_{\,q}\,(q,\,u)\,, \\
  u^{\,\prime}\ &=\ 2\,(u\ -\ c)^2\,\Bb\,(q,\,u)\ \defeq\ \F_{\,u}\,(q,\,u)\,.
\end{align*}
The equilibria points might be of two kinds:
\begin{enumerate}
  \item on the intersection of curves $\Gamma_{\,\Aa}\ \eqdef\ \bigl\{\Aa\,(q,\,u)\ =\ 0\bigr\}$ and $\Gamma_{\,\Bb}\ \eqdef\ \bigl\{\Bb\,(q,\,u)\ =\ 0\bigr\}\,$,
  \item on the intersection of $\bigl\{\Dd\,(u)\ =\ 0\bigr\}$ and $\bigl\{\Bb\,(q,\,u)\ =\ 0\bigr\}\,$.
\end{enumerate}
The linear stability of equilibria is determined by eigenvalues of the following \textsc{Jacobian} matrix:
\begin{equation}\label{eq:eigs}
  \Jj\,(q,\,u;\,\delta,\,c)\ \eqdef\ \begin{pmatrix}
    \displaystyle{\pd{\F_{\,q}}{q}} & \displaystyle{\pd{\F_{\,q}}{u}} \vspace*{2mm} \\
    \displaystyle{\pd{\F_{\,u}}{q}} & \displaystyle{\pd{\F_{\,u}}{u}}
  \end{pmatrix}\,.
\end{equation}
The elements of the \textsc{Jacobian} matrix can be computed by direct differentiation:
\begin{align*}
  \Jj_{1\,1}\ &=\ -2\,\sigma\,(1\ +\ \delta)\,q\,\Dd\,(u)\,, \\
  \Jj_{1\,2}\ &=\ \sigma\,\biggl[\,\Bigl[\,2\,u\ -\ \frac{\M}{(u\ -\ c)^2}\,\Bigr]\ +\ \Bigl[\,2\,(u\ -\ c)\ -\ \frac{\M}{(u\ -\ c)^2}\,\Bigr]\,\Aa\,(q,\,u)\,\biggr]\,, \\
  \Jj_{2\,1}\ &=\ -\sigma\,(u\ -\ c)^2\,\Bigl[\,2\,u\ -\ \frac{\M}{(u\ -\ c)^2}\,\Bigr]\,, \\
  \Jj_{2\,2}\ &=\ 4\,(u\ -\ c)\,\Bb\,(q,\,u)\ +\ 2\,\bigl(\alpha\,\M\ -\ \sigma\,q\,\Dd\,(u)\bigr)\,.
\end{align*}
The last expression can be simplified at equilibria locations by taking into account the fact that $\Aa\ =\ \Bb\ \equiv\ 0$ at the first kind and $\Dd\ =\ \Bb\ \equiv\ 0$ at equilibria states of the second kind.

\begin{figure}
  \centering
  \subfigure[]{\includegraphics[width=0.47\textwidth]{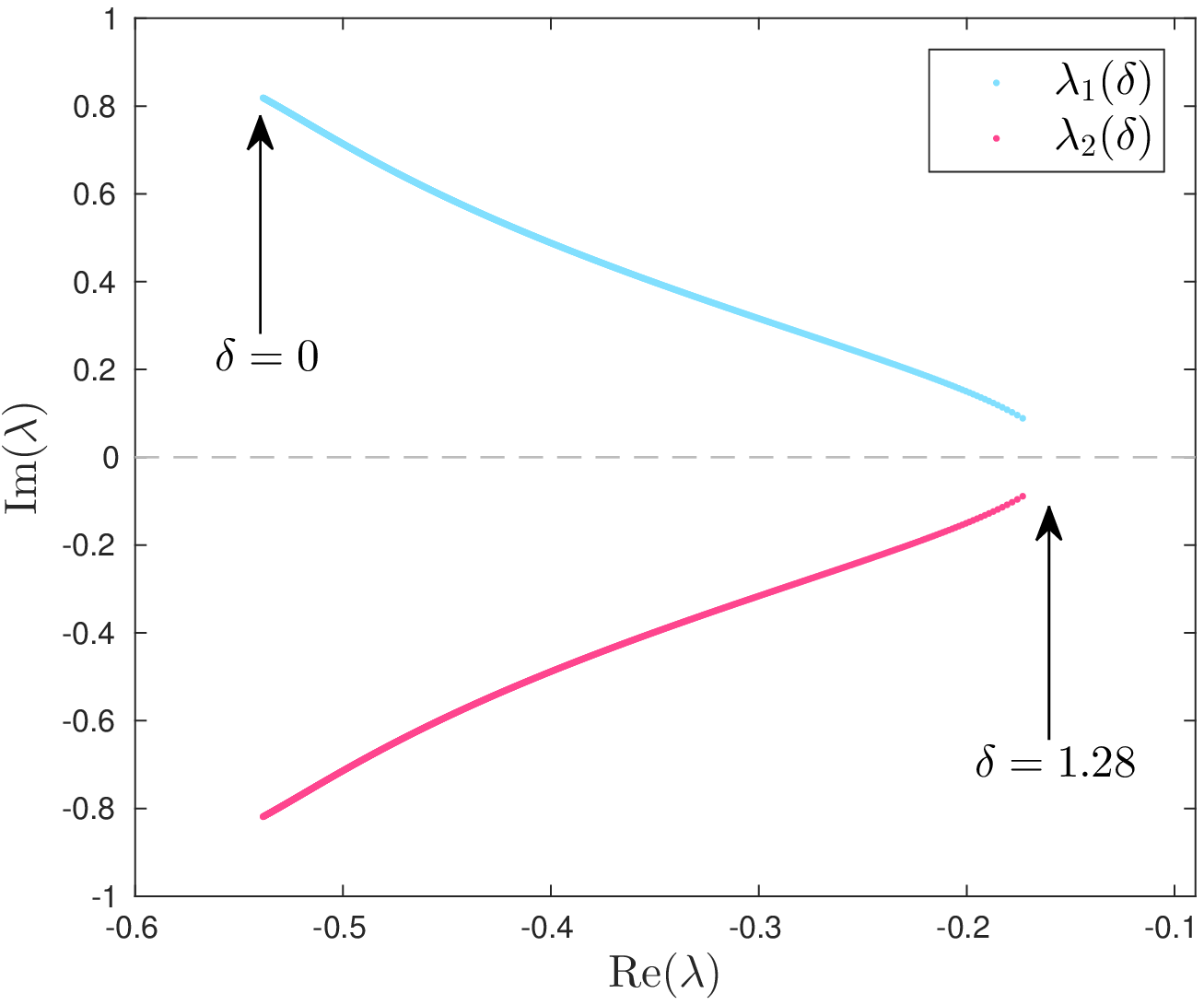}}
  \subfigure[]{\includegraphics[width=0.51\textwidth]{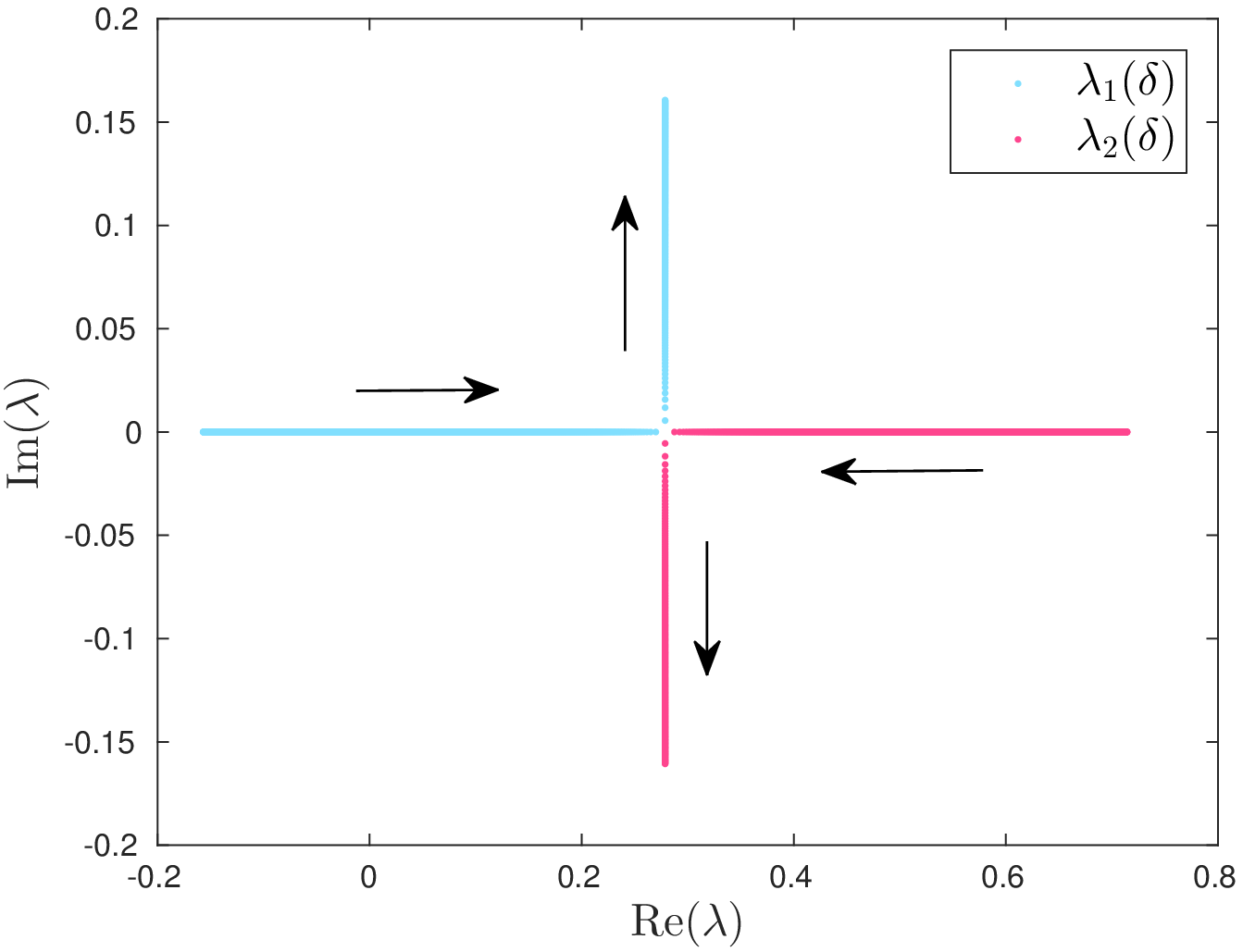}}
  \caption{\small\em Eigenvalues of the \textsc{Jacobian} \eqref{eq:eigs} for various values of the parameter $\delta\ \in\ [\,0,\,1.28\,]\,$. The left and right panels (a, b) correspond to equilibria of the first and second kinds correspondingly. Eigenvalue positive real part indicates the instability of the equilibrium point \cite{Lyapunov1892}. The arrows in the right panel indicate the increasing direction of the parameter $\delta\,$.}
  \label{fig:eigs}
\end{figure}

Eigenvalues $\lambda_{1,\,2}\,(\delta)$ of the \textsc{Jacobian} matrix $\Jj$ as functions of the parameter $\delta$ for both types of equilibria points are shown in Figure~\ref{fig:eigs}. The physical parameters used in this numerical computation are $\sigma\ =\ 0.15\,$, $\alpha\ =\ \tan\phi\ =\ 1.0\,$, $c\ =\ 2.787$ and $\M\ =\ 0.1\,c\,$. From this illustration it is clear now that, at least for these values of parameters, the equilibria of the first kind are unstable spiral points, while the equilibria of the second kind are (linearly) stable spirals \cite{Lyapunov1892}.

%%% ----------------------------------------------------------------------- %%%

\subsubsection{A particular class of travelling waves}
\label{sec:TW}

Similarly to the construction of steady solutions presented in the previous Section~\ref{sec:steady}, we are looking for \emph{travelling wave solutions} to system \eqref{eq:eq1} -- \eqref{eq:eq4} of the form:
\begin{equation*}
  h\,(t,\,x)\ =\ h\,(\xi)\,, \quad u\,(t,\,x)\ =\ u\,(\xi)\,, \quad m\,(t,\,x)\ =\ m\,(\xi)\,,
\end{equation*}
where $\xi\ \equiv\ x\ -\ c\cdot t$ is a combined independent variable introduced earlier. In other words, we consider the frame of reference where the travelling wave is steady. Notice that the model \eqref{eq:eq1} -- \eqref{eq:eq4} does not possess the \textsc{Galilean} invariance property because of the velocity variables present in the right hand sides. This effect comes from the assumption that the ambient fluid remains in rest, which privileges this particular frame of reference. The constant $c\ >\ 0$ is the unknown wave celerity to be determined during the solution procedure.

Consider travelling waves of the following particular form:
\begin{equation*}
  u\,(\xi)\ =\ u_{\,f}\,, \quad m\,(\xi)\ =\ m_{\,f}\,, \quad q\,(\xi)\ =\ q_{\,f}\,, \quad h\,(\xi)\ =\ h_{\,f}\ -\ \varsigma\,\xi\,,
\end{equation*}
with $\varsigma\ >\ 0$ and $\xi\ <\ 0\,$. The travelling wave ansatz presented above satisfies the following relations:
\begin{align}\label{eq:6a}
  \varsigma\cdot\bigl(c\ -\ u_{\,f}\bigr)\ =\ \sigma\cdot q_{\,f}\,, \\
  \varsigma\cdot\bigl((c\ -\ u_{\,f})\,u_{\,f}\ -\ \half\,m_{\,f}\bigr)\ =\ \alpha\cdot m_{\,f}\,, \\
  \varsigma\cdot\bigl((c\ -\ u_{\,f})\cdot \bigl(u_{\,f}^{\,2}\ +\ q_{\,f}^{\,2}\ +\ m_{\,f}\bigr)\ -\ m_{\,f}\cdot u_{\,f}\bigr)\ =\ 2\,\alpha\,m_{\,f}\cdot u_{\,f}\ -\ \kappa\cdot q_{\,f}^{\,3}\,.\label{eq:6c}
\end{align}
We use also an additional hypothesis that the flow in the coordinate frame moving with the travelling wave is critical, \ie
\begin{equation}\label{eq:7}
  \bigl(u_{\,f}\ -\ c\bigr)^{\,2}\ =\ m_{\,f}\,.
\end{equation}
The last condition can be equivalently recast as
\begin{equation*}
  \Fr_{\,f}\ \eqdef\ \frac{\abs{c\ -\ u_{\,f}}}{\sqrt{m_{\,f}}}\ \equiv\ 1\,,
\end{equation*}
where $\Fr_{\,f}$ is the \textsc{Froude} number with respect to the wave front \cite{Fer2002a}. This condition ensures the existence of the self-sustained regime of the wave propagation independently of small perturbations which might occur behind the wave front. It is completely analogous to the so-called \textsc{Chapman}--\textsc{Jouguet} condition for the propagation of a self-sustained detonation wave in gas dynamics \cite{Zeldovich1940}.

\begin{figure}
  \centering
  \includegraphics[width=0.89\textwidth]{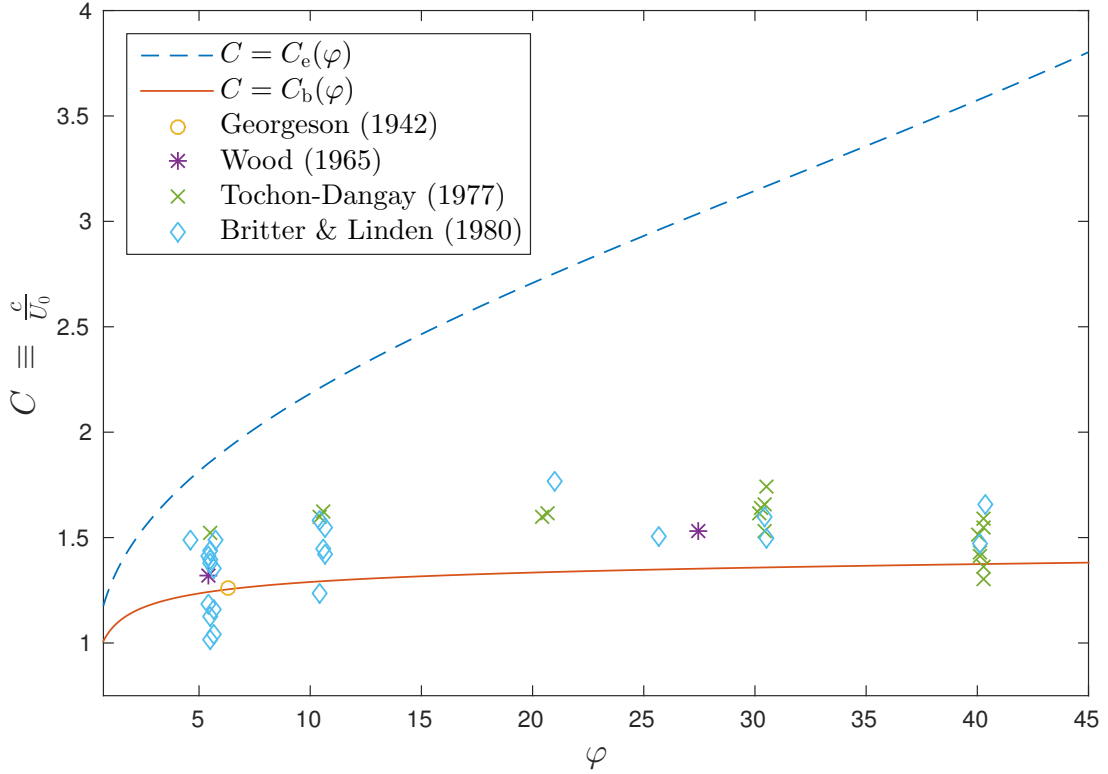}
  \caption{\small\em Dependence of the flow head speed on the slope angle: analytical predictions against experimental measurements \cite{Georgeson1942, Wood1965, Tochon-Dangay1977, Britter1980}. The friction parameter used is $c_{\,w}\ =\ 0.004\,$. The velocity profile given by the dashed line $C\ =\ C_{\,\mathrm{e}}\,(\varphi)$ is determined by relations \eqref{eq:6a} -- \eqref{eq:6c}, while the profile represented by the solid line $C\ =\ C_{\,\mathrm{b}}\,(\varphi)$ is defined by relations \eqref{eq:bc1} -- \eqref{eq:bc3}.}
  \label{fig:britten}
\end{figure}

The wave celerity $c$ might be eliminated from Equations \eqref{eq:6a} -- \eqref{eq:7} by introducing new variables:
\begin{equation*}
  u^{\,\star}\ \eqdef\ \frac{u_{\,f}}{c}\,, \qquad
  q^{\,\star}\ \eqdef\ \frac{q_{\,f}}{c}\,, \qquad
  m^{\,\star}\ \eqdef\ \frac{m_{\,f}}{c^2}\,.
\end{equation*}
As a result, we come to the following closed system of equations:
\begin{align}\label{eq:8a}
  \varsigma\cdot(1\ -\ u^{\,\star})\ &=\ \sigma\cdot q^{\,\star}\,, \\
  \varsigma\cdot(1\ -\ u^{\,\star})\,u^{\,\star}\ -\ \half\;\varsigma\cdot\,m^{\,\star}\ &=\ \alpha\cdot m^{\,\star}\,, \label{eq:8b} \\
  \varsigma\cdot\bigl((u^{\,\star})^2\ +\ (q^{\,\star})^2\ +\ m^{\,\star}\bigr)\ -\ \varsigma\cdot m^{\,\star}\cdot u^{\,\star}\ &=\ 2\,\alpha\,m^{\,\star}\cdot u^{\,\star}\ -\ \kappa\cdot(q^{\,\star})^3\,,\label{eq:8c} \\
  \bigl(1\ -\ u^{\,\star}\bigr)^2\ &=\ m^{\,\star}\,.\label{eq:8d}
\end{align}
By assuming that $\sigma\cdot q^{\,\star}\ \neq\ 0\,$, from last equations we can derive a simple relation:
\begin{equation*}
  (u^{\,\star})^2\ -\ m^{\,\star}\ -\ (1\ +\ \delta)\cdot(q^{\,\star})^2\ =\ 0\,,
\end{equation*}
and by using the scaled version \eqref{eq:8d} of relation \eqref{eq:7}, we obtain that
\begin{equation*}
  2\,u^{\,\star}\ -\ 1\ =\ (1\ +\ \delta)\cdot(q^{\,\star})^2\,,
\end{equation*}
or formally,
\begin{equation*}
  q^{\,\star}\ =\ \sqrt{\frac{2\,u^{\,\star}\ -\ 1}{1\ +\ \delta}}\,,
\end{equation*}
provided that $u^{\,\star}\ >\ \half$ so that the value $q^{\,\star}\ \in\ \R^{\,+}\,$. After dividing \eqref{eq:8b} by \eqref{eq:8a} we have:
\begin{equation*}
  \frac{(1\ -\ u^{\,\star})\,u^{\,\star}\ -\ \half\;m^{\,\star}}{1\ -\ u^{\,\star}}\ \equiv\ \frac{1}{2}\;\bigl(3\,u^{\,\star}\ -\ 1\bigr)\ =\ \frac{\alpha\,\sqrt{1\ +\ \delta}\,(1\ -\ u^{\,\star})^2}{\sigma\,\sqrt{2\,u^{\,\star}\ -\ 1}}\,.
\end{equation*}
In other words, we have the following equation for $u^{\,\star}\,$:
\begin{equation}\label{eq:k}
  3\,u^{\,\star}\ -\ 1\ =\ \beta\;\frac{(1\ -\ u^{\,\star})^2}{\sqrt{2\,u^{\,\star}\ -\ 1}}\,, \qquad \mathrm{ with }\qquad \beta\ \eqdef\ \frac{2\,\alpha\,\sqrt{1\ +\ \delta}}{\sigma}\,.
\end{equation}
Under the same condition on $u^{\,\star}\,$, we can transform Equation~\eqref{eq:k} into the following algebraic equation:
\begin{equation*}
  (3\,u^{\,\star}\ -\ 1)^2\cdot\bigl(2\,u^{\,\star}\ -\ 1\bigr)\ =\ \beta^{\,2}\cdot(1\ -\ u^{\,\star})^4\,.
\end{equation*}
It can be shown (see~\ref{app:tw}) that there exists a unique positive root to this equation, which belongs to the interval $u^{\,\star}\ \in\ (\,\half,\,1\,)\,$. All other remaining quantities such as $\varsigma\,$, $m^{\,\star}$ and $q^{\,\star}$ are determined after finding $u^{\,\star}\,$. The velocity $c$ can be found from the mass conservation equation written at the wave front:
\begin{equation*}
  (c\ -\ w_{\,b})\cdot m_{\,b}\ +\ (c\ -\ u_{\,f})\cdot m_{\,f}\ =\ c\cdot m_{\,s}\,, \qquad w_{\,b}\ >\ c\,.
\end{equation*}
If $m_{\,s}\ \equiv\ 0\,$, the last equation takes a particularly simple form:
\begin{equation*}
  \frac{(c\ -\ w_{\,b})^3}{(U_{\,b})^3}\ +\ \bigl(\Fr_{\,b}\bigr)^{\,-\frac{2}{3}}\;\frac{c}{U_{\,b}}\ -\ 1\ =\ 0\,,
\end{equation*}
where we introduced some notations:
\begin{equation*}
  U_{\,b}\ \eqdef\ \sqrt[3]{\M_{\,b}}\,, \qquad
  \Fr_{\,b}\ \eqdef\ \frac{u_{\,b}}{\sqrt{m_{\,b}}}\ =\ \sqrt{\frac{\alpha}{c_{\,w}}}\,.
\end{equation*}
We remind that the positive constant $c_{\,w}$ controls the magnitude of bottom friction effects. We thus arrive to the following cubic polynomial equation for the quantity $C\ \eqdef\ \dfrac{c}{U_{\,b}}\,$:
\begin{equation*}
  \P\,(C)\ \equiv\ (1\ -\ u^{\,\star})^3\,C^{\,3}\ +\ \bigl(\Fr_{\,b}\bigr)^{\,-\frac{2}{3}}\,C\ -\ 1\ =\ 0\,.
\end{equation*}
It is not difficult to see that the polynomial $\P(C)$ is a monotonically increasing function of $C$ provided that $u^{\,\star}\ \leq\ 1$ and, thus, there exists a unique positive root (since $\P\,(0)\ = \ -1\ <\ 0$). The dependence of the computed in this way coefficient $C\ =\ C_{\,e}\,(\Fr_{\,b})\ =\ C_{\,e}\,(\phi)$ on the channel slope angle $\phi$ is shown in Figure~\ref{fig:britten} with the dashed line (for a fixed value of $c_{\,w}\ =\ 0.004$ and with $\sigma\ =\ 0.15\,$, $\delta\ =\ 4\,$). Note that in the experimental study \cite{Britter1980} the notation $C\ \equiv\ \dfrac{c}{U_{\,0}}$ was used, while we consider another definition $C\ \equiv\ \dfrac{c}{U_{\,b}}\,$, where $U_{\,0}\ \eqdef\ \sqrt[3]{\M_{\,0}}\,$. We can see that this prediction does not compare very well with the experimental data \cite{Georgeson1942, Wood1965, Tochon-Dangay1977, Britter1980} for large slope angles $\phi\,$. This drawback will be corrected below. The agreement for small angles of inclination is achieved since $\M_{\,j}\ \ll\ \M_{\,b}$ and, consequently, $U_{\,0}\ -\ U_{\,b}\ \ll\ U_{\,b}\,$. Therefore, the total mass flux $\M_{\,0}$ can be replaced by the bottom flux $\M_{\b}$ to find the density current head velocity. It is also worth to mention that the big scatter in experimental data depicted in Figure~\ref{fig:britten} can be also partly explained by the differences between the total $\M_{\,0}$ and bottom $\M_{\,b}$ mass fluxes, which depend on precise inflow conditions.

%%% ----------------------------------------------------------------------- %%%

\subsubsection{Further considerations}

The stability of the constructed travelling wave solution with the linear growth of the profile (due to the perpetual entrainment of sediments and still water into the mixing layer) has to be studied separately. The constructed solution belongs to the important class of stratified flows with fluid particles (in some parts of this flow) moving faster than the wave front. It is known that for surface waves it leads to inevitable wave breaking \cite{Massel2007}. The authors are not aware of any mathematical stability studies of such stratified flows.

In the modelling of density currents, where there are heavy particles (of density $\rho_{\,0}$) entrained into the flow, the ``boundary'' conditions imposed on the wave front have the capital importance. In the previous Section these conditions were determined by the flow structure and we saw that it leads to the poor prediction of the wave celerity comparing with experimental data \cite{Georgeson1942, Wood1965, Tochon-Dangay1977, Britter1980}. Here we propose another set of ``boundary'' conditions:
\begin{align}\label{eq:bc1}
  m\cdot(c\ -\ u)\ +\ b_{\,0}\,\zeta_{\,b}\cdot(c\ -\ w_{\,b})\ &=\ 0\,, \\
  h\cdot\bigl((u\ -\ c)\,u\ +\ \half\;m)\bigr)\ &=\ 0\,,\label{eq:bc2} \\
  \bigl(c\ -\ u\bigr)^2\ &=\ m\,,\label{eq:bc3}
\end{align}
which are composed of the mass conservation equation, the flow criticality condition relative to the wave front and also the momentum conservation equation. It is not difficult to see that for $u\ <\ c\,$, it follows that $u\ =\ \third\;c\,$ and for the dimensionless combination $C\ =\ \frac{c}{U_{\,b}}$ we obtain the following cubic equation:
\begin{equation}\label{eq:stst}
  \frac{8}{27}\;C^{\,3}\ +\ \Fr_{\,b}^{\,-\frac{2}{3}}\,C\ -\ 1\ =\ 0\,.
\end{equation}
All intermediate computations are explained in~\ref{app:der}. For large values of the \textsc{Froude} number, the solution $C\ \approx\ 1.5\,$. This estimation is in good agreement with the empirical conjecture on the density current flow head velocity based on the experiments reported in \cite{Britter1980}:
\begin{equation*}
  c\ \approx\ 1.5\times \M_{\,0}^{\,\frac{1}{3}}\,.
\end{equation*}
In Figure~\ref{fig:britten} the solid line shows the dependence of the flow head celerity $C\ =\ C_{\,b}\,(\phi)$ based on Equation~\eqref{eq:stst}. The overall good agreement of the lower solid curve with the experimental data \cite{Georgeson1942, Wood1965, Tochon-Dangay1977, Britter1980} certifies the model quality. It is worth to notice that the proportionality coefficient in experimental studies was determined based on velocity $\M_{\,0}^{\,\frac{1}{3}}$ and not on velocity $U_{\,b}\,$. As a result, for more accurate comparisons with considered models, one has to determine accurately, which part of the mass flow $\M_{\,0}$ entered into the boundary layer, since $U_{\,b}\ =\ \M_{\,b}^{\,\frac{1}{3}}\ =\ \Bigl(\mu\,\M_{\,0}\Bigr)^{\,\frac{1}{3}}\,$, with $\mu\ \leq\ 1\,$.

%%% ----------------------------------------------------------------------- %%%

\subsection{Similarity solutions}
\label{sec:sim}

Additionally to special solutions considered in two previous Sections, System \eqref{eq:eq1} -- \eqref{eq:eq4} admits also the following class self-similar solutions:
\begin{multline}\label{eq:star}
  h\,(x,\,t)\ =\ t^{\,\uptheta+1}\,\hat{h}\,(\xi)\,, \qquad
  u\,(x,\,t)\ =\ t^{\,\uptheta}\,\hat{u}\,(\xi)\,, \\
  m\,(x,\,t)\ =\ t^{\,2\,\uptheta}\,\hat{m}\,(\xi)\,, \qquad
  q\,(x,\,t)\ =\ t^{\,\uptheta}\,\hat{q}\,(\xi)\,,
\end{multline}
where $\xi\ \eqdef\ \dfrac{x}{t^{\,\uptheta+1}}$ and $\uptheta\ \in\ \R\,$. For flows with the sustained mass flow $\M_{\,b}\ \equiv\ \const$ or with the sustained sediment mass $m_{\,s}\ \equiv\ \const\,$, from conditions on the front it follows that $\uptheta\ \equiv\ 0\,$. For the sustained mass flow in the turbulent mixing layer $\M_{\,j}\ \equiv\ \const\,$, the same value for $\uptheta$ follows from boundary conditions, \ie $\uptheta\ \equiv\ 0\,$. Henceforth, for all just mentioned cases, we look for solutions of the form:
\begin{equation}\label{eq:twostars}
  h\,(x,\,t)\ =\ t\cdot\hat{h}\,(\xi)\,, \qquad
  u\,(x,\,t)\ =\ \hat{u}\,(\xi)\,, \qquad
  m\,(x,\,t)\ =\ \hat{m}\,(\xi)\,, \qquad
  q\,(x,\,t)\ =\ \hat{q}\,(\xi)\,,
\end{equation}
with $\xi\ \eqdef\ \dfrac{x}{t}\,$. Such solutions play a very important r\^ole in understanding the system \eqref{eq:eq1} -- \eqref{eq:eq4}, since dynamic solutions tend to self-similar ones at large times provided that constant mass fluxes are maintained.

Another important class of self-similar solutions \eqref{eq:star} is realized when $\M_{\,0}\ \equiv\ 0\,$, $m_{\,s}\ \equiv\ 0\,$. This corresponds to the evolution of a finite mass of a heavier liquid, which propagates under the layer of a lighter fluid. In this case, the mass conservation law
\begin{equation*}
  \int_{\,0}^{\,+\infty} m\,(t,\,x)\;\ud x\ =\ \const
\end{equation*}
yields self-similar solutions of the form \eqref{eq:star}:
\begin{multline*}
  h\,(x,\,t)\, =\, t^{\,2/3}\,\hat{h}\,(\xi)\,, \qquad
  u\,(x,\,t)\, =\, t^{\,-1/3}\,\hat{u}\,(\xi)\,, \\
  m\,(x,\,t)\, =\, t^{\,-2/3}\,\hat{m}\,(\xi)\,, \qquad
  q\,(x,\,t)\, =\, t^{\,-1/3}\,\hat{q}\,(\xi)\,,
\end{multline*}
where $\xi\ \eqdef\ \dfrac{x}{t^{\,2/3}}\,$. This self-similar solution indicates \cite{Ross2006} that the wave front position $x_{\,f}$ behaves asymptotically in time as:
\begin{equation*}
  x_{\,f}\ \propto\ t^{\,2/3}\,.
\end{equation*}
From the last estimation, the following asymptotic behaviour of the front velocity $U_{\,f}$ can be readily deduced:
\begin{equation}\label{eq:ass}
  U_{\,f}\ \propto\ x_{\,f}^{-1/2}\,.
\end{equation}
The last decay law was validated experimentally in \cite{Beghin1981, Maxworthy2010}. In the present work this asymptotic behaviour will be used to validate the numerical simulations. This approach to the description of turbidity currents is referred as the ``\emph{thermal theory}'' \cite{Beghin1981}.

In turbidity flows, in order to construct self-similar solutions for density currents along a slope (as well as for travelling waves), it is of capital importance to prescribe the \emph{adequate} boundary conditions. During the construction of travelling waves above, we see that the wave celerity depends on the relations imposed at the wave front. When the solution is given by ansatz \eqref{eq:star}, the flow head is uniquely determined. However, the applicability of such similarity solutions to flows in the absence of sediments (\ie $m_{\,s}\ \equiv\ 0\,$, $\M_{\,b}\ \equiv\ 0$) has to be studied separately.

The situation changes when we consider the problem of sediments entrainment into the flow ($m_{\,s}\ >\ 0\,$, $\M_{\,b}\ \equiv\ 0$). In the framework of the conservative system \eqref{eq:cons1} -- \eqref{eq:cons4} this problem can be interpreted as the mixed Initial--Boundary Value Problem (IBVP). Namely, at the initial moment of time we know the distribution of heavy sediments in the sediment layer $h_{\,s}\,(x)\,$, $m_{\,s}\,(x)\,$. The sediments might be at rest ($u_{\,s}\,(x)\ \equiv\ 0$) or moving with prescribed velocity $u_{\,s}\,(x)$ and turbulent kinetic energy $q_{\,s}\,(x)\,$. The perturbations entraining the sediments into the flow are entering the fluid domain from the left boundary (without any loss of generality). In Figure~\ref{fig:sketch} we depicted the sketch of the fluid domain for an illustration. In this situation there is no need to separate the flow head and to write additional relations on the wave front, since the main flow characteristics are obtained in the process of solving the IBVP. However, the main question remains unanswered: which self-similar regime \eqref{eq:star} will appear as the long time limit of the unsteady solution?

System \eqref{eq:cons1} -- \eqref{eq:cons4} is of hyperbolic (and hydrodynamic) type and it corresponds to the flow of a barotropic gas with chemical reactions in compressible fluid dynamics. We have already mentioned above the analogy between density currents and the detonation theory. During the normal detonation, the fluid flow satisfying the \textsc{Chapman}--\textsc{Jouguet} conditions downstream the flow head is not always realized (especially in the presence of accompanying chemical reactions). Under certain conditions the wave of detonation might propagate with the velocity exceeding that of perturbations behind the wave front \cite{Fickett1979}. The main conclusion that we can draw from this analogy is that the realizability of self-similar solutions has to be studied separately. Such an analytical study might turn out to be very complex. Below we will show by numerical means that the propagation of the flow head with the speed higher than expected is possible. We would like to mention also that similarity solutions for gravity currents were constructed also in \cite{Ross2006} and \cite[Sections~3 \& 4]{Johnson2013a}.

%%% ----------------------------------------------------------------------- %%%

\section{Model validation and unsteady simulations}
\label{sec:exp}

Strictly speaking, we already validated steady solutions to System \eqref{eq:steady1} -- \eqref{eq:steady6} by making comparisons with the experiments from \cite{Pawlak2000}. We show that this system is able to predict qualitatively and quantitatively the development of the mixing layer over a slope.

%%% ----------------------------------------------------------------------- %%%

\subsection{Problem formulation}

In this Section we continue the validations by considering unsteady solutions hereinafter. Moreover, we shall consider the applicability of an even simpler one-layer model: 
\begin{align}\label{eq:psi1}
  h_t\ +\ [\,h\,u\,]_x\ &=\ \chi^+\,, \\
  u_t\ +\ u\,u_x\ +\ b\,h_x\ +\ \half\,h\,b_x\ &=\ -b\,d_x\ -\ b_0\,\zeta_x\ -\ \frac{\psi\,\chi^-\ +\ u\,\chi^+}{h}\,, \\
  b_t\ +\ u\,b_x\ &=\ -\,\frac{b_{\,0}\,\chi^{\,-}\ +\ b\,\chi^{\,+}}{h}\,, \\
  q_t\ +\ u\,q_x\ &=\ (2\,q\,h)^{-1}\;\bigl[\,(2\,\psi\,u\ -\ \psi^2\ +\ b_0\,h\ -\ 2\,b\,h)\,\chi^- \nonumber \\
  &\ +\ (u^2\ -\ q^2\ -\ b\,h)\,\chi^+\ -\ \kappa\,q^3\,\bigr]\,.\label{eq:psi2}
\end{align}
We apply it to simulate the sediments entrainment process by density currents over moderate (finite) slopes. Numerical solutions will be compared with experimental data reported in \cite{Rastello2004} as well as with exact special solutions of travelling wave type. In their experiments the dense fluid consisted either of saltwater or of the sawdust particle suspension. The shape of these particles was irregular in accordance with suspended snow flakes. The spatial growth of the cloud was determined from the side view images recorded with a video camera. A $5\ \cm$ square grid was drawn on the side glass to facilitate the front position determination. Moreover, we shall show below that under certain initial conditions the numerical solution will tend asymptotically to self-similar solutions described earlier.

We made our choice for experimental data of \textsc{Rastello} \& \textsc{Hopfinger} (2004) \cite{Rastello2004} for the following reasons:
\begin{itemize}
  \item This experiment corresponds fairly well to the scope and purpose of our numerical model
  \item The model \eqref{eq:psi1} -- \eqref{eq:psi2} is suitable for the simulation of gravity currents even over moderate and large slopes used in the experimental study \cite{Rastello2004} (see Table~\ref{tab:params} for the values of the slope parameter). When using other models, one has to check that the influence of the sediment bottom layer is taken into account to represent correctly the front dynamics
  \item The experiment was conducted for sufficiently long time. In this way we are able to check our model validity for the acceleration and deceleration stages of the flow. Finally, we were even able to check the asymptotic behaviour of the flow, which is very typical for buoyant flows over a slope.
\end{itemize}
 
\begin{figure}
  \centering
  \includegraphics[width=0.99\textwidth]{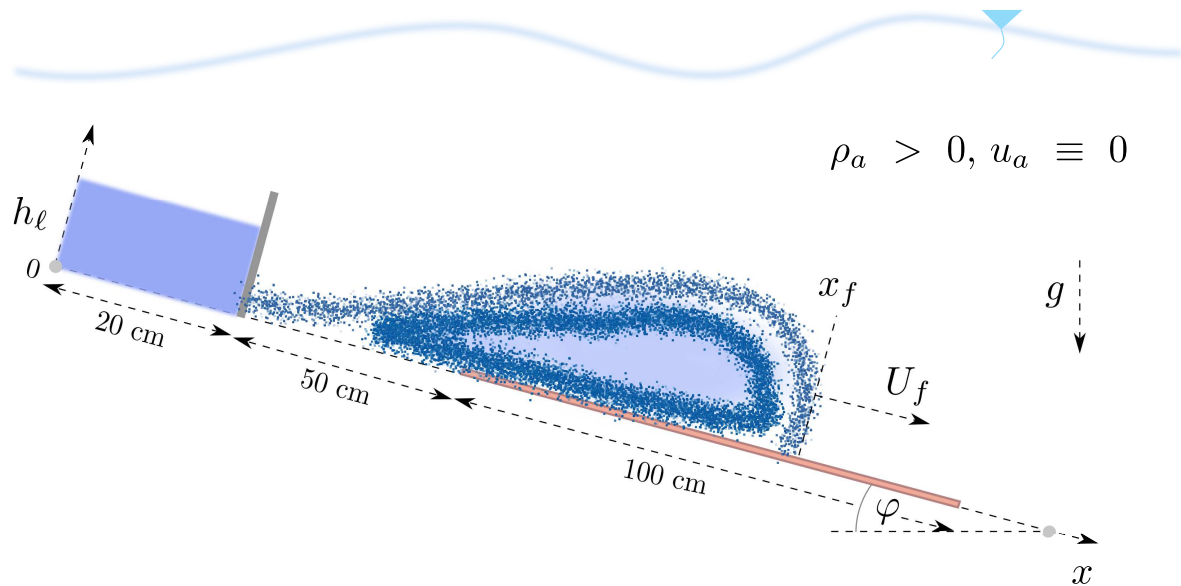}
  \caption{\small\em Sketch of numerical lock-exchange experiments.}
  \label{fig:expSketch}
\end{figure}

We already mentioned above that the equilibrium system \eqref{eq:psi1} -- \eqref{eq:psi2} possesses the hyperbolic structure similar to the equations of gas dynamics with two ``sonic'' and two contact characteristics \cite{Rozhdestvenskiy1978, Godlewski1990}. There is a wide choice for the numerical discretization of such systems. However, taking into account that we deal with a system of conservation (balance) laws, it seems to be natural to opt for finite volume schemes \cite{Barth2004}. Henceforth, in order to solve numerically the equilibrium system \eqref{eq:psi1} -- \eqref{eq:psi2} we use the classical and robust finite volume discretization and the widely used \textsc{Godunov} scheme \cite{Godunov1959, Godunov1987, Godunov1999}. The explicit \textsc{Euler} scheme is used in time discretization.

Sketch of the numerical experiment is given in Figure~\ref{fig:expSketch}. Our goal is to reproduce \emph{in silico} some of the laboratory experiments reported in \cite{Rastello2004}. The values of all physical parameters are given in Table~\ref{tab:params}. The initial and boundary conditions are rather standard as well. Initially, at $t\ =\ 0\,$, the distribution of evolutionary quantities is given on  the computational domain $[\,0,\,\ell\,]\,$:
\begin{equation*}
  h\,(x,\,0)\ =\ h_{\,0}\,(x)\,, \quad
  u\,(x,\,0)\ =\ u_{\,0}\,(x)\,, \quad
  m\,(x,\,0)\ =\ m_{\,0}\,(x)\,, \quad
  q\,(x,\,0)\ =\ q_{\,0}\,(x)\,.
\end{equation*}
On the right boundary we set wall boundary conditions for simplicity\footnote{In any case, the simulation stops before the mass reaches the right boundary. So, the influence of the right boundary condition on presented numerical results is completely negligible.}. On the left extremity of the computational domain the boundary conditions depend on the flow regime in the vicinity of the left boundary. If the flow there is supercritical, we impose \textsc{Cauchy}'s data. Otherwise, we impose a wall boundary condition as well. The case we study below is depicted in Figure~\ref{fig:expSketch} and it will be described in more details below.

%%% ----------------------------------------------------------------------- %%%

\subsection{Experimental set-up}

In order to reproduce \emph{in silico} the experiments of \textsc{Rastello} \& \textsc{Hopfinger} (2004) \cite{Rastello2004} we use the following configuration of the numerical tank. The channel length $\ell$ is equal to $200\;\cm\,$. As we already mentioned, on boundaries we impose wall boundary conditions, in other words the channel is closed as in experiments. A heavier fluid of buoyancy $b_{\,\ell}\ \in\ \bigl\{1,\,19\bigr\}\;\dfrac{\cm}{\s^{\,2}}$ fills an initially closed container of the length of $20\;\cm$ and of variable height $h_{\,\ell}\,$, which changed from one experiment to another. This configuration corresponds to the classical lock-exchange experiment. The heavy fluid ``mass\footnote{We employ the term mass in the sense of the relative weight of the dense fluid.}'' $m_{\,\ell}\ =\ b_{\,\ell}\cdot h_{\,\ell}\,$. At the distance of $50\;\cm$ from this recipient the slope is covered by initially motionless sediment layer of the height $\zeta_{\,s}\ =\ 0.2\;\cm$ and $2\;\cm\,$. The ``mass'' of sediments is $m_{\,s}\ =\ b_{\,s}\cdot \zeta_{\,s}\,$. The length of sediments layer is $100\;\cm\,$. During the propagation of the heavy fluid head all these sediments were entrained into the flow. In some experiments the sediment layer was absent, \ie $m_{\,s}\ \equiv\ 0\,$. In this case the flow simply propagates over the rigid inclined bottom. In order to avoid the degeneration of certain equations, in numerical experiments we cover the whole slope with a micro-layer of sediments with mass $m_{\,s}^{\,\circ}\ >\ 0\,$, such that
\begin{equation*}
  m_{\,s}^{\,\circ}\ \ll\ m_{\,s} \qquad \mbox{and} \qquad
  m_{\,s}^{\,\circ}\ \ll\ m_{\,\ell}\,.
\end{equation*}
The values of all physical parameters are given in Table~\ref{tab:params}. The influence of the model parameters on predicted values was found to be rather weak. Consequently, in all numerical simulations reported below we used the following values of these parameters:
\begin{equation*}
  \sigma\ =\ 0.15\,, \qquad
  \delta\ =\ 0\,, \qquad
  c_{\,w}\ =\ 0\,.
\end{equation*}
The results of the critical comparisons with laboratory data are discussed below.

\begin{table}
  \centering
  \begin{tabular}{c|l|l}
  \hline\hline
  \textit{Parameter} & \textit{Experiment 1} & \textit{Experiment 2} \\
  \hline\hline
  Slope angle, $\varphi$ & $32^\circ$ & $45^\circ$ \\
  Container height, $h_{\,\ell}$ & $6.5\ \cm$ & $20\ \cm$ \\
  Heavy fluid buoyancy, $b_{\,\ell}$ & $19\ \frac{\cm}{s^2}$ & $1\ \frac{\cm}{s^2}$ \vspace*{0.2em} \\
  Sediment deposit height, $\zeta_{\,s}$ & $0.2\ \cm$ & $2.0\ \cm$ \\
  Minimal sediment ``mass'', $m_{\,s}^{\,\circ}$ & $10^{-3}\;\frac{\cm^2}{\s^2}$ & $10^{-3}\;\frac{\cm^2}{\s^2}$ \vspace*{0.2em} \\
  Sediment ``mass'', $m_{\,s}$ & $10\;\frac{\cm^2}{\s^2}$; $20\;\frac{\cm^2}{\s^2}$ & $3\;\frac{\cm^2}{\s^2}$ \vspace*{0.2em} \\
  Final simulation time, $T$ & $10\ \s$ & $20\ \s$ \\
  \hline\hline
  \end{tabular}
  \bigskip
  \caption{\small\em Parameters used in numerical simulations of the experiments from \textsc{Rastello} \& \textsc{Hopfinger} (2004) \cite[Table~2]{Rastello2004}, schematically depicted in Figure~\ref{fig:expSketch}.}
  \label{tab:params}
\end{table}

%%% ----------------------------------------------------------------------- %%%

\subsection{Numerical results}

\begin{figure}
  \centering
  \subfigure[$\phi\ =\ 32^\circ$]{\includegraphics[width=0.5\textwidth]{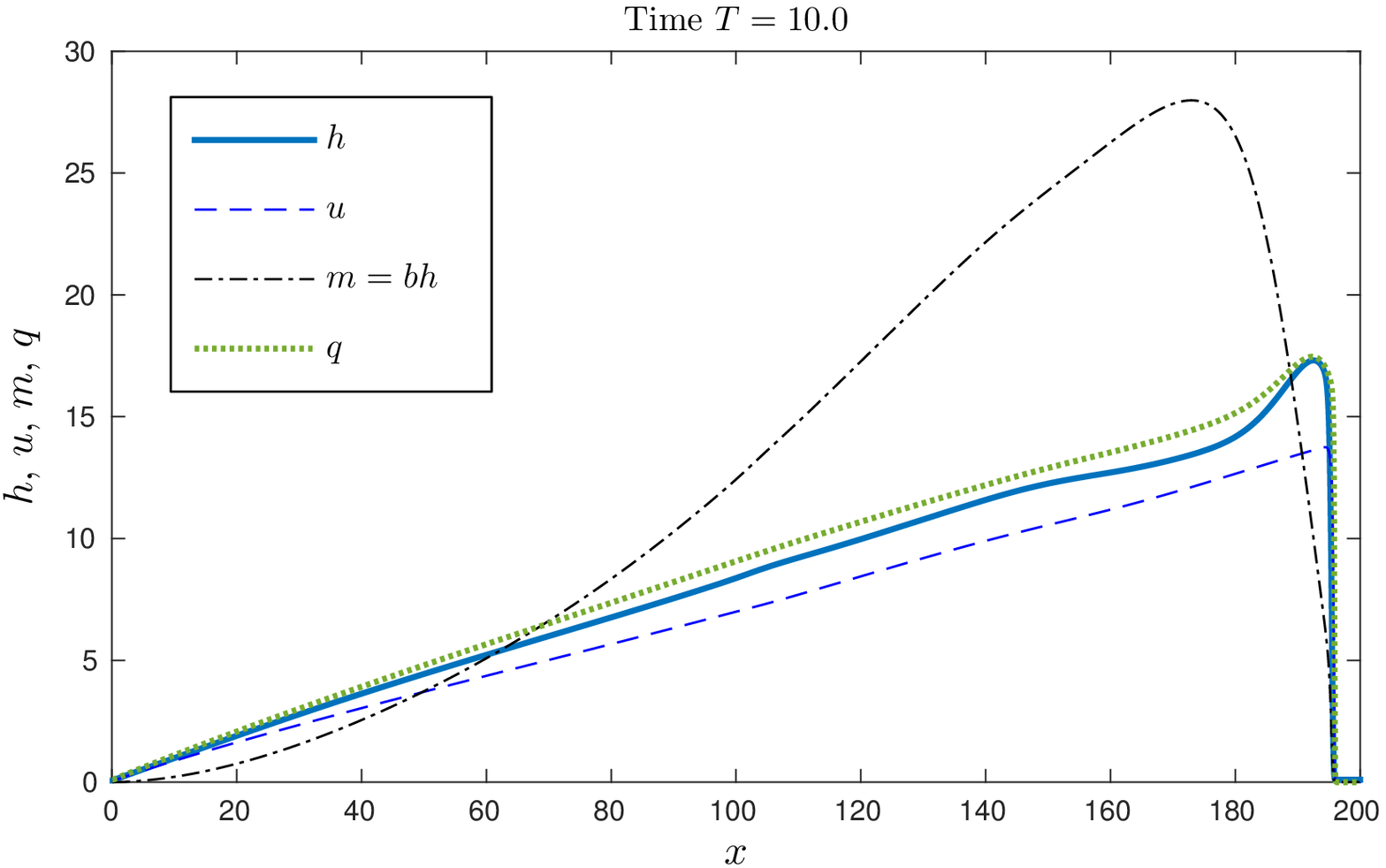}}
  \subfigure[$\phi\ =\ 45^\circ$]{\includegraphics[width=0.45\textwidth]{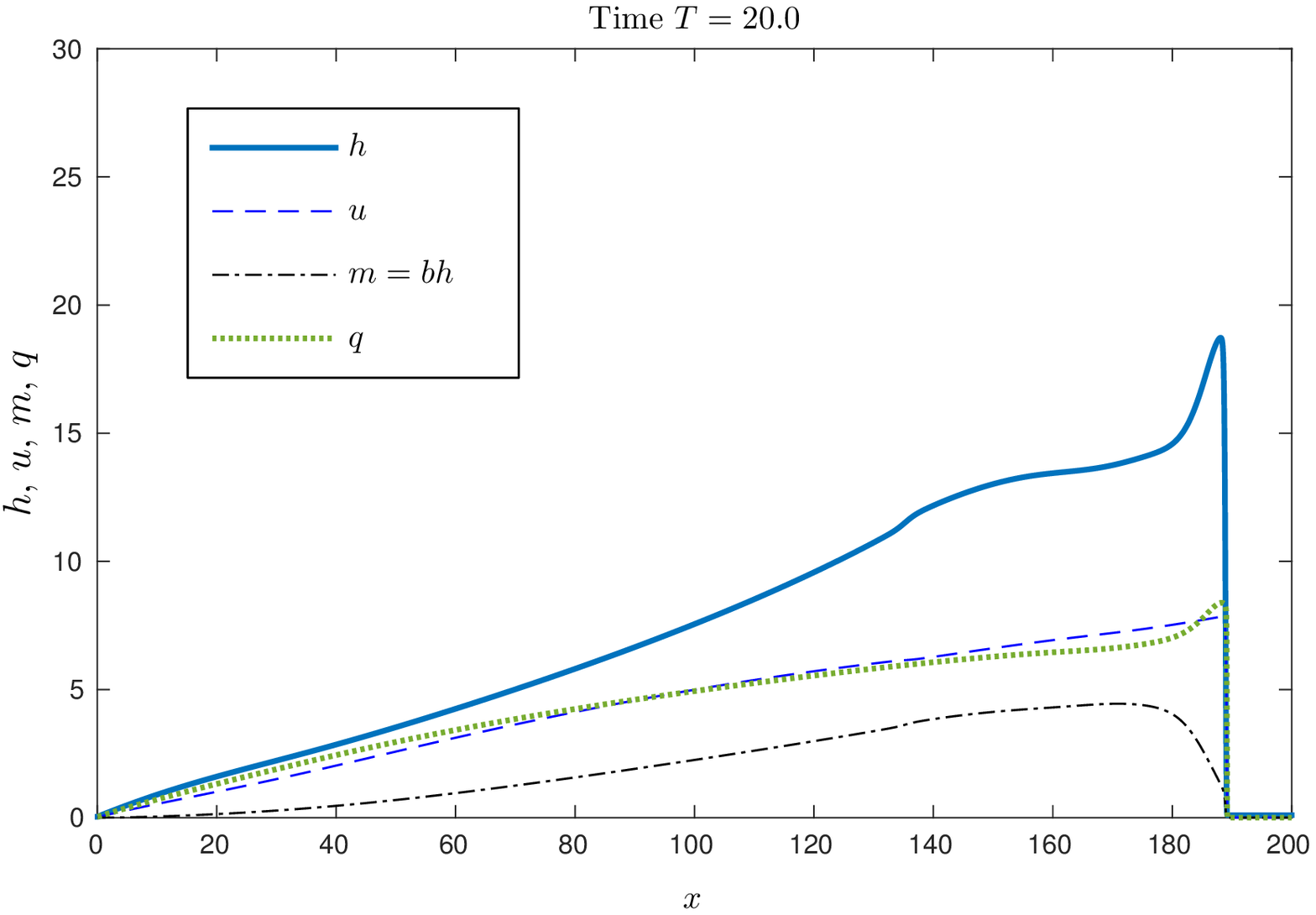}}
  \caption{\small\em Distribution of depth-integrated physical quantities $h\,$, $u\,$, $m$ and $q$ in numerical computations reproducing the experimental set-up from \cite{Rastello2004}. These simulations are performed without the presence of sediments along the slope.}
  \label{fig:distrib}
\end{figure}

In Figure~\ref{fig:distrib} we report the results of numerical simulations showing the spatial profiles of four quantities $h\,$, $u\,$, $m$ and $q\,$. These simulations were performed without the sediments layer, \ie $m_{\,s}\ \equiv\ 0\,$. The panels correspond to:
\begin{description}
  \item[Left panel] Slope angle $\phi\ =\ 32^{\circ}$, the snapshot is taken at $T\ =\ 10\;\s$
  \item[Right panel] Slope angle $\phi\ =\ 45^{\circ}$, the snapshot is taken at $T\ =\ 20\;\s\,$.
\end{description}
All quantities reported in Figure~\ref{fig:distrib} are given in dimensional variables. The position of the density current front is clearly visible in both cases despite the fact that the IBVP was solved in the domain $[\,0,\,\ell\,]\times[\,0,\,T\,]$ using simple shock-capturing methods (\ie no special treatment was necessary to detect the front). This property of the employed numerical scheme will be used to analyze below the asymptotic behaviour of unsteady solutions.

\begin{landscape}
\begin{figure}
  \centering
  \includegraphics[width=1.45\textwidth]{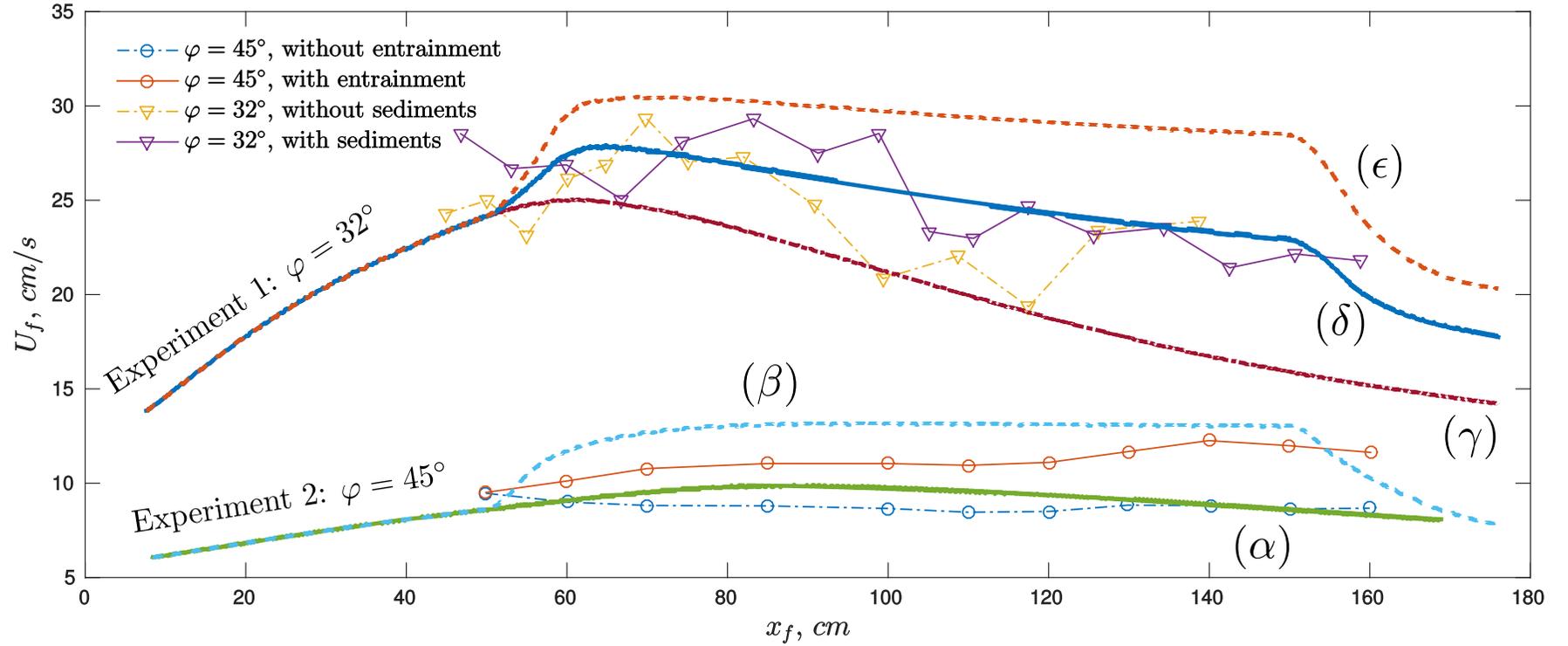}
  \caption{\small\em Evolution of the front velocity $U_{\,f}$ of laboratory clouds as a function of downstream distance $x_{\,f}\,$. Comparison of the predicted flow head velocity (lines without markers) against the experimental results (lines with markers) from Rastello \& Hopfinger (2004) \cite[Fig.~11]{Rastello2004}. The upper (lower) group of data corresponds to Experiment~1 (2) from Table~\ref{tab:params} correspondingly: $(\alpha)$ $m_{\,s}\ =\ 10^{-3}\ \textsf{cm}^{\,2}/\textsf{s}^{\,2}\,$; $(\beta)$ $m_{\,s}\ =\ 3\ \textsf{cm}^{\,2}/\textsf{s}^{\,2}\,$; $(\gamma)$ $m_{\,s}\ =\ 10^{-3}\ \textsf{cm}^{\,2}/\textsf{s}^{\,2}\,$; $(\delta)$ $m_{\,s}\ =\ 10\ \textsf{cm}^{\,2}/\textsf{s}^{\,2}\,$; $(\epsilon)$ $m_{\,s}\ =\ 20\ \textsf{cm}^{\,2}/\textsf{s}^{\,2}\,$.}
  \label{fig:hopfComp}
\end{figure}
\end{landscape}

In Figure~\ref{fig:hopfComp} we show the dependence of the flow head velocity $U_{\,f}$ on the distance $x_{\,f}$ traveled by the front. The parameters of these numerical/laboratory Experiments~1 \& 2 are given in Table~\ref{tab:params}. Please, note that the heavy fluid density in Experiment~1 is much higher. The last observation explains why the head velocity in Experiments~1 is higher than in Experiments~2, even if the slope is bigger in Experiments~$2\,$. Various symbols ($\bigcirc$ and $\bigtriangledown$) correspond to laboratory measurements of the head velocity and are taken from \cite[Fig.~11]{Rastello2004}. Solid and dash-dotted lines (with $\bigcirc$ and $\bigtriangledown$) indicate the experiments with ($m_{\,s}\ >\ 0$) and without ($m_{\,s}\ =\ 0$) sediments correspondingly. The presence of sediments before the head front yields the initial acceleration of the flow. This acceleration phase is followed by deceleration, since sediments are distributed over a limited distance in our experiments. A very good agreement with numerical predictions can be observed in Figure~\ref{fig:hopfComp}.

\begin{figure}
  \centering
  \includegraphics[width=0.89\textwidth]{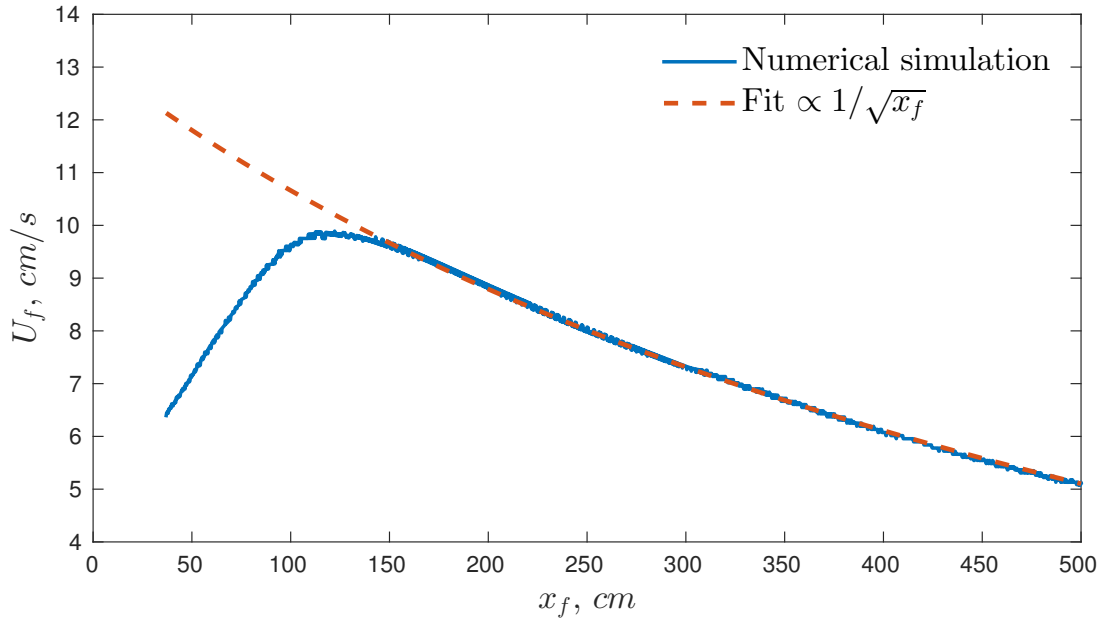}
  \caption{\small\em Asymptotic behaviour \eqref{eq:ass} of the front velocity for the experiment $2$ from Table~\ref{tab:params}.}
  \label{fig:asympt}
\end{figure}

In Figure~\ref{fig:asympt} we show the long time behaviour of the front velocity from the Experiment~$2$ reported in Figure~\ref{fig:hopfComp} with parameters from Table~\ref{tab:params}. In order to perform this simulation we increased the computational domain\footnote{This operation can be made easily in numerical experiments contrary to laboratory experiments, where the channel size is rather fixed.} from $200$ to $500\;\cm\,$. Sediments were absent along the slope in this computation (\ie $m_{\,s}\ \equiv\ 0$). The excellent agreement with theoretically established asymptotics $U_{\,f}\ \sim\ \dfrac{1}{\sqrt{x_{\,f}}}$ can be observed in this Figure~\ref{fig:asympt}. It validates both the numerics and the underlying theoretical argument.

%%% ----------------------------------------------------------------------- %%%

\section{Discussion}
\label{sec:disc}

Above we have proposed and tested a new model for density currents. We outline below the main conclusions and perspectives of this study.

%%% ----------------------------------------------------------------------- %%%

\subsection{Conclusions}

In the present study we considered the problem of density current modelling propagating down the slope in the presence of sediment deposits on the rigid bottom. In order to describe this flow in mathematical terms, we divide it into three zones (from the bottom vertically upwards): the layer of sediments (i), the mixing layer (ii) and the still water layer (iii). In order to simplify the mathematical description, we assumed that the upper layer (iii) of the still water was motionless. The depth-averaged description was adopted in each of two remaining layers (i) and (ii). As a result, we arrived to a shallow water two-layer system including turbulent modelling, which can be recast in a conservative form \eqref{eq:nc1} -- \eqref{eq:nc6} of quasi-linear balance laws \cite{Rozhdestvenskiy1978}. The initially proposed system contains only six evolution equations to describe the complex density current. However, it can be further simplified if we assume that the flow in the bottom layer occupied by dense liquid or suspension with uplifted sediments is in the equilibrium state, \ie the gravity force is exactly balanced by the friction forces with the rigid bottom. In this way, we can remove two equations corresponding to the bottom layer (i). The sediment equilibrium model is given by System \eqref{eq:psi1} -- \eqref{eq:psi2}. This concludes the modelling part of our study.

Then, the proposed equilibrium model is studied using analytical means. Namely, we were interested in some special, but very important classes of solutions --- steady states, travelling waves and similarity flows. The linear stability of travelling waves was studied. The initial conditions which yield asymptotically (in time) some special solutions (such as self-similar and travelling waves) were discussed. The velocity of travelling waves is of capital importance for the understanding of turbidity flows propagation. Namely, the travelling wave speed gives an estimation of the flow head propagation along the slope. This prediction of our travelling wave analysis was checked against the experimental data from numerous previous studies \cite{Georgeson1942, Wood1965, Tochon-Dangay1977, Britter1980}.

The mixing layer formation was discussed in the framework of steady solutions and our theoretical prediction was compared against experimental data from \cite{Pawlak1998}. Moreover, unsteady solutions predicted by our model were compared with density flow experiments made in the LEGI (UMR 5519) Laboratory by \textsc{Rastello} \& \textsc{Hopfinger} (2004) \cite{Rastello2004}. The model predictions are in good agreement with their measurements of the front velocity. Finally, the asymptotic behaviour for the flow head velocity as a function of the front position was measured in our numerical computations and an excellent agreement was found with existing theoretical estimations. This concludes the validation of the model and also of several particular solutions derived in this study.

It is important to underline that the proposed model describes three different flow regimes:
\begin{enumerate}
  \item Exchange flow,
  \item Thermals,
  \item Underwater avalanches.
\end{enumerate}
In the first case, the permanent (or transient) inflow of a heavy fluid mass accelerates downslope. In natural environments this situation can be realized when a heavy fluid overflows a bottom obstacle or a hill, which creates a deposit of the heavy fluid mass behind the obstacle. Our mathematical model showed that the initial accelerating phase in the development of the supercritical flow is of capital importance since it yields the formation of the so-called mixing layer. In particular, the density current head velocity depends directly on the portion of the mass entering the bottom layer. It is this sea-bed underflow, which supplies the head directly almost without mixing. This fact may explain also the relatively important scattering of observed front velocities in laboratory experiments (see \eg Figure~\ref{fig:britten}).

In order to describe the evolution of thermals over a downslope, one can use the simplified model \eqref{eq:cons1} -- \eqref{eq:cons4}, which is a direct generalization of usual nonlinear shallow water equations taking into account the ambient quiescent fluid entrainment into the flow. It is important to notice also that the entrainment speed in our model is determined automatically together with other flow parameters. Figure~\ref{fig:asympt} showed also that this model can be used to describe thermal flows during the acceleration \emph{and} deceleration phases simultaneously.

Density currents over a sloping bottom featuring sediments entrainment into the flow have numerous applications in geophysics such as snow and underwater avalanches, seasonal mass exchanges in deep lakes, \etc These flows are situated in the center of different mathematical modelling efforts mentioned in Introduction. Unfortunately, contrary to the previous two situations (lock-exchange and thermal flows), laboratory experiments and real-world measurements still remain seldom. It is in this field of applications that the r\^ole of simple mathematical models increases substantially. Such models should allow to predict theoretically at least the front velocity and, if possible, several other parameters of the self-sustained current. In the particular situation where we neglect the sedimentation velocity, the one layer model \eqref{eq:cons1} -- \eqref{eq:cons4} is suitable for the description of underwater avalanches. Moreover, our computation showed that the flow achieves quickly a self-similar regime, which will be studied in future publications. However, in contrast to exchange flows, the condition \eqref{eq:bc3} of the flow criticality behind the wave front (analogous to the \textsc{Chapman}--\textsc{Jouguet} condition in detonation) is not fulfilled in the presence of sediments entrainment. It is another flow regime, which is realized in practice: the flow turns out to be substantially supercritical in the frame of reference moving with the front. This situation is usually referred to as ``\emph{overdriven detonation}'', when the main flow parameters, such as detonation pressure and propagating velocity, exceed the corresponding \textsc{Chapman}--\textsc{Jouguet} values \cite{Liu2000}. A further study is necessary to understand the front velocity selection mechanism for such flows. In this way we come naturally to the description of some other perspectives of our study.

%%% ----------------------------------------------------------------------- %%%

\subsection{Perspectives}

In the present study, we have employed only basic legacy first order \textsc{Godunov} finite volume schemes \cite{Godunov1959, Godunov1987, Godunov1999}. Their principal advantages are the ease of implementations and the robustness of numerical results. However, the accuracy might be also important in some applications, where a quantitative prediction is critically important. Consequently, some high order finite volume well-balanced schemes have to be developed to solve numerically the density current models proposed in our study. This technology is relatively well mastered nowadays \cite{Dutykh2010e, Dutykh2013a, Dutykh2013, Gosse2013}. 

The system of equations presented earlier in this article is a simplified equilibrium model based on a number of physical assumptions. Of course, more complete models of higher physical fidelity should be developed as well in the future along the lines outlined in the monograph \cite{Liapidevskii2000}. For instance, in this manuscript we considered density currents with uniform cross-sections, \ie 2D flows. In future works 3D effects have to be included. Moreover, in all examples considered earlier, the uniform bed slopes was used. In part, it comes from the experimental set-up used in previous investigations. However, in upcoming works the interaction between the fluid flow with the bed morphology has to be investigated to advance our understanding of these processes.

%%% ----------------------------------------------------------------------- %%%

\subsection*{Acknowledgments}
\addcontentsline{toc}{subsection}{Acknowledgments}

This research was supported by the project \textsc{Aval} (\textsc{AAP Montagne 2015}, University Savoie Mont Blanc) and Russian Science Foundation (project 15-11-20013). V.~\textsc{Liapidevskii} acknowledges the CNRS/ASR Cooperation Program under the project \textnumero 23975 and the hospitality of the University Savoie Mont Blanc during his numerous visits. The authors are grateful to Professors L.~\textsc{Armi} \& G.~\textsc{Pawlak} for their help in preparing this manuscript.

%%% ----------------------------------------------------------------------- %%%

\appendix
\section{An exact solution}
\label{app:exact}

In the Appendix we present the construction of a particular steady solution. Using the information presented above in Sections~\ref{sec:steady} and \ref{sec:travel}, we construct in this Section the profile of a particular exact solution to Equations \eqref{eq:eq1} -- \eqref{eq:eq4}. We assume that in the bottom layer $\zeta\ \equiv\ \zeta_{\,b}$ and $w\ \equiv\ w_{\,b}\,$. Moreover, we suppose that the \emph{total} mass inflow $\M_{\,0}\ =\ b_{\,0}\,\zeta_{\,0}\,w_{\,0}$ is prescribed on the left boundary. This incident mass flux $\M_{\,0}$ splits into two streams:
\begin{equation*}
  \M_{\,0}\ =\ \M_{\,j}\ +\ \M_{\,b}\,, \qquad
  \M_{\,j}\ \eqdef\ \frac{1}{2}\;b_{\,0}\,h_{\,L}\,u_{\,L}\,, \qquad
  \M_{\,b}\ \eqdef\ b_{\,0}\,\zeta_{\,b}\,w_{\,b}\,.
\end{equation*}
We assume also that the ratio $\mu\ \eqdef\ \dfrac{\M_{\,j}}{\M_{\,0}}\ \in\ (0,\,1\,]$ is fixed. Thus, $\M_{\,b}$ can be also easily deduced.

The flow at any fixed moment of time $t\ =\ t_{\,m}\ >\ 0$ is composed of two parts:
\begin{enumerate}
  \item \textbf{Stationary flow:} $u\,(x,\,t)\ \equiv\ u_{\,j}\,$, $m\,(x,\,t)\ \equiv\ m_{\,j}\,$, $q\,(x,\,t)\ \equiv\ q_{\,j}\,$ and $h\,(x,\,t)\ =\ h_{\,j}\ +\ \varsigma_{\,j}\cdot x\,$, where $x\ \in\ [\,0,\,L_{\,j}\,]\,$, $L_{\,j}\ =\ u_{\,j}\cdot t_{\,m}\,$.
  \item \textbf{Travelling wave:} This part depends on the variable $\xi\ =\ x\ -\ c\cdot t$ and $u\,(\xi)\ \equiv\ u_{\,f}\,$, $m\,(\xi)\ \equiv\ m_{\,f}\,$, $q\,(\xi)\ \equiv\ q_{\,f}\,$ and $h\,(\xi)\ =\ h_{\,f}\ -\ \varsigma_{\,f}\cdot(x\ -\ L_{\,t})\,$, where $\varsigma_{\,f}\ >\ 0\,$, $L_{\,t}\ -\ L_{\,f}\ \leq\ x\ \leq\ L_{\,t}$ and $L_{\,f}\ \eqdef\ u_{\,f}\cdot t_{\,m}\,$, $L_{\,t}\ \eqdef\ c\cdot t_{\,m}\,$.
\end{enumerate}
The problem consists in finding all solution parameters. The solution algorithm is summarized below.

%%% ----------------------------------------------------------------------- %%%

\subsection{Steady flow}
\label{app:A1}

Let $a\ \eqdef\ \dfrac{u_{\,j}}{U_{\,j}}$ be the ratio of two speeds with $U_{\,j}\ =\ \sqrt[3]{\M_{\,j}}$ and $z\ \eqdef\ a^3\,$. Then, one has to find the unique root $z\ >\ 1$ to the following equation:
\begin{equation*}
  \Bigl(1\ -\ \frac{1}{z}\Bigr)\cdot(1\ +\ 2\,z)^2\ =\ \frac{4\,\alpha^2\,(1\ +\ \delta)}{\sigma^2}\ \defeq\ \beta^2\,,
\end{equation*}
which is to be compared with Equation~\eqref{eq:k}. It is equivalent to finding the root to the following cubic polynomial:
\begin{equation*}
  \P_{\,1}\,(z)\ =\ 4\,z^3\ -\ (3\ +\ \beta^2)\,z\ -\ 1\,.
\end{equation*}
Let $z^{\star}$ be the required root. Then, we find $a\ =\ \sqrt[3]{z^{\star}}$ and $u_{\,j}\ =\ a\,U_{\,j}\,$. We know that $\M_{\,j}\ =\ U_{\,j}^{\,3}\ =\ m_{\,j}\,u_{\,j}\,$. Hence, $m_{\,j}\ =\ \dfrac{U_{\,j}^{\,2}}{a}\,$. The remaining variables are:
\begin{equation*}
  q_{\,j}\ =\ \sqrt{\frac{u_{\,j}^{\,2}\ -\ m_{\,j}}{1\ +\ \delta}}\,, \qquad
  \varsigma_{\,j}\ =\ \frac{\sigma\cdot q_{\,j}}{u_{\,j}}\,, \qquad
  b_{\,L}\ =\ \frac{b_{\,0}}{2}\,, \qquad
  h_{\,L}\ =\ \frac{m_{\,j}}{b_{\,L}}\,.
\end{equation*}
Finally, the steady part of the turbulent jet layer depth is equal to
\begin{equation*}
  h\,(x,\,t)\ =\ h_{\,L}\ +\ \varsigma_{\,j}\cdot x\,, \qquad
  0\ \leq\ x\ \leq\ L_{\,j}\ =\ u_{\,j}\cdot t_{\,m}\,.
\end{equation*}

%%% ----------------------------------------------------------------------- %%%

\subsection{Travelling waves}
\label{app:tw}

\begin{figure}
  \centering
  \subfigure[$\mu\ =\ 0.05$]{\includegraphics[width=0.48\textwidth]{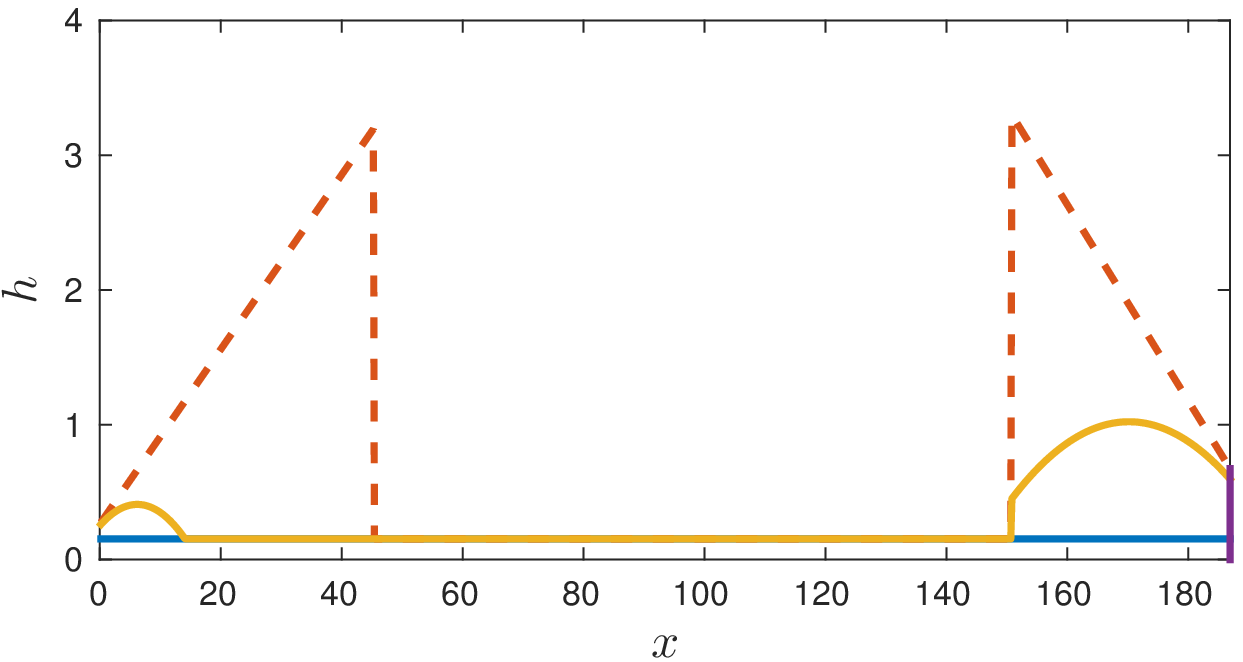}}
  \subfigure[$\mu\ =\ 0.654$]{\includegraphics[width=0.48\textwidth]{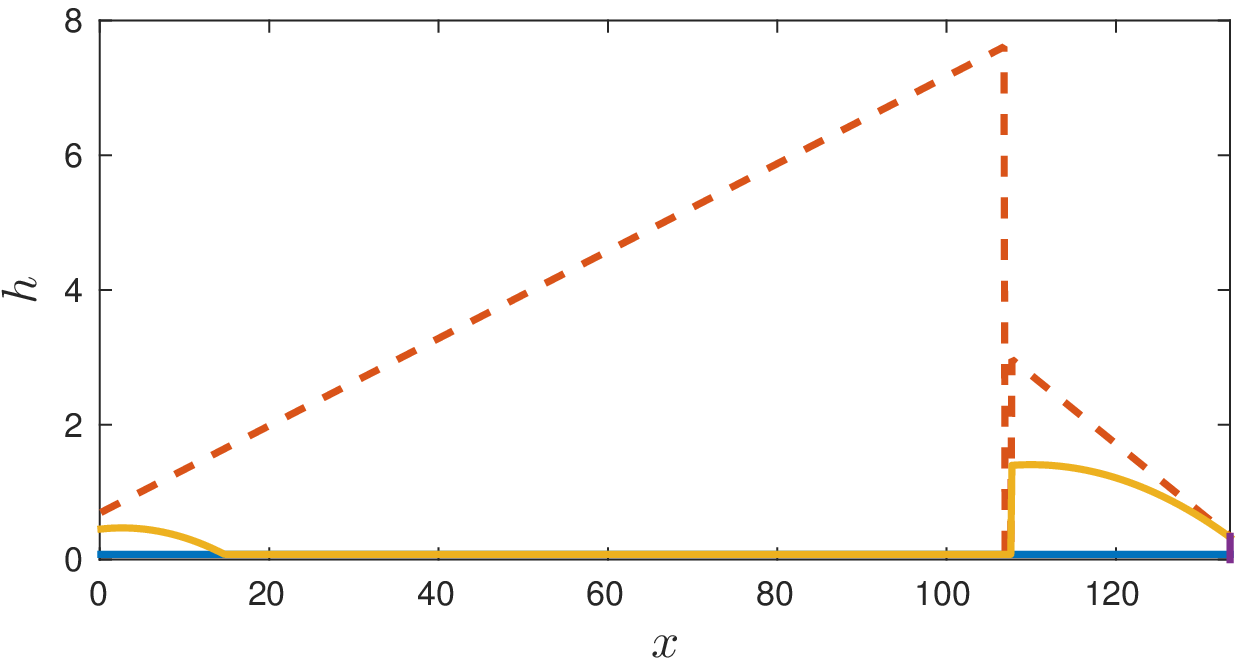}}
  \caption{\small\em Analytical solutions depicted at $t_{\,m}\ =\ 50\ \s\,$. All parameters are the same except for (a) $\mu\ =\ 0.05$ and (b) $\mu\ =\ 0.654\,$. The exact solutions are represented with dashed lines, while visible boundaries are shown with solid lines. Solutions are terminated at the right boundary by the vertical sharp front. Notice different vertical scales on left and right panels.}
  \label{fig:anal}
\end{figure}

For the travelling wave part we have $\M_{\,b}\ =\ U_{\,b}^{\,3}$ and $\Fr_{\,b}\ \eqdef\ \dfrac{w_{\,b}}{\sqrt{b_{\,0}\,\zeta_{\,b}}}\,$. Let $u^{\,\star}\ \eqdef\ \frac{u_{\,f}}{c}$ be the dimensionless flow velocity. It can be found by solving the following algebraic equation, which possesses a real root on the interval $u^{\,\star}\ \in\ \bigl(\half,\,1\bigr)\,$:
\begin{equation*}
  \P_{\,2}\,(u^{\,\star})\ =\ \beta^{\,2}\,(1\ -\ u^{\,\star})^4\ -\ (2\,u^{\,\star}\ -\ 1)\,(3\,u^{\,\star}\ -\ 1)^2\ =\ 0\,,
\end{equation*}
where the constant $\beta$ was defined earlier in \eqref{eq:k}. The polynomial $\P_{\,2}\,(u^{\,\star})$ can be expanded:
\begin{equation*}
  \P_{\,2}\,(u^{\,\star})\ =\ (u^{\,\star})^4\ -\ (4\ +\ 18\,\beta^{\,-2})\,(u^{\,\star})^3\ +\ (6\ +\ 21\,\beta^{\,-2})\,(u^{\,\star})^2\ -\ (4\ +\ 8\,\beta^{\,-2})\,u^{\,\star}\ +\ 1\ +\ \beta^{\,-2}\,.
\end{equation*}
Let us assume that we have found the required root $u^{\,\star}\ \in\ \bigl(\half,\,1\bigr)\,$. Then, we solve another polynomial equation to determine $C\ =\ \frac{c}{U_{\,b}}\,$:
\begin{equation*}
  \P_{\,3}\,(C)\ =\ (1\ -\ u^{\,\star})^3\,C^{\,3}\ +\ \Fr_{\,b}^{\,-2/3}\,C\ -\ 1\ =\ 0\,.
\end{equation*}
The last equation admits a unique positive root $C^{\,\star}\ =\ C^{\,\star}\,(\Fr_{\,b})\ =\ C_{\,e}\,(\phi)$ as well. The last dependence is shown in Figure~\ref{fig:britten} with the dashed line. Then, we consider the flow in the bottom layer. We have the following relations:
\begin{equation*}
  \M_{\,b}\ =\ b_{\,0}\,\zeta_{\,b}\,w_{\,b}\ \equiv\ U_{\,b}^{\,3}\,, \qquad
  w_{\,b}\ =\ \Fr_{\,b}\,\sqrt{b_{\,0}\,\zeta_{\,b}}\ \equiv\ \Fr_{\,b}\,\sqrt{m_{\,b}}\,,
\end{equation*}
\begin{equation*}
  m_{\,b}\ =\ \Fr_{\,b}^{\,-2/3}\cdot \M_{\,b}^{\,2/3}\,, \qquad
  \zeta_{\,b}\ =\ \frac{m_{\,b}}{b_{\,0}}\,.
\end{equation*}
We can notice that $w_{\,b}$ can be also written as $w_{\,b}\ =\ \Fr_{\,b}^{\,2/3}\,U_{\,b}\,$. Now we come back to the travelling wave:
\begin{equation*}
  c\ =\ C^{\,\star}\,U_{\,b}\,, \qquad
  u_{\,f}\ =\ u^{\,\star}\,c\,,
\end{equation*}
and from the relation
\begin{equation*}
  (c\ -\ u_{\,f})\cdot b_{\,0}\,h_{\,f}\ =\ (w_{\,b}\ -\ c)\cdot m_{\,b}
\end{equation*}
we can find
\begin{equation*}
  h_{\,f}\ =\ \frac{(w_{\,b}\ -\ c)\cdot m_{\,b}}{(c\ -\ u_{\,f})\cdot b_{\,0}}\,,
\end{equation*}
provided that $(c\ -\ u_{\,f})\cdot b_{\,0}\ \neq\ 0\,$. Finally, we find two last elements of the solution:
\begin{equation*}
  q_{\,f}\ =\ \sqrt{\frac{2\,\dfrac{u_{\,f}}{c}\ -\ 1}{1\ +\ \delta}}\,, \qquad
  \varsigma_{\,f}\ =\ \frac{\sigma\,q_{\,f}}{c\ -\ u_{\,f}}\,.
\end{equation*}
The travelling wave profile is then given by
\begin{equation*}
  h\,(\xi)\ =\ h_{\,f}\ -\ \varsigma_{\,f}\cdot(x\ -\ L_{\,t})\,, \qquad
  L_{\,t}\ =\ c\cdot t_{\,m}\,, \qquad L_{\,f}\ =\ u_{\,f}\cdot t_{\,m}\,.
\end{equation*}
The last profile is located in the segment $x\ \in\ [\,L_{\,t}\ -\ L_{\,f},\,L_{\,t}\,]\,$. The solution exists if $u_{\,j}\ \leq\ u_{\,f}\,$. We underline that in the region between the stationary and travelling wave parts the flow in general is unsteady. Thus, the indicated boundaries $L_{\,t}\ -\ L_{\,f}$ and $L_{\,t}$ are rather approximations to the reality. Two such exact solutions for two different values of the parameter $\mu\ =\ 0.05$ (left panel) and $\mu\ =\ 0.654$ (right panel) are depicted in Figure~\ref{fig:anal}. This picture shows that at least two different configurations can be realized in practice:
\begin{description}
  \item[Two-wave (left panel (\textit{a}))] when the main portion of the heavy fluid enters into the boundary layer ($\mu\ =\ 0.05$)
  \item[One-wave (right panel (\textit{b}))] when the flow in the buoyant jet reaches the head front ($\mu\ =\ 0.654$).
\end{description}

\begin{remark}
In experiments in order to highlight the processes happening in the flow, the heavy fluid of density $\rho_{\,0}$ is in general coloured. Other experimental visualization techniques are also available, but the fluid painting is the most widely used one. By $\rho_{\,a}$ we denote the density of the light ambient fluid. In such experimental conditions the \emph{visible} interface between two fluids is located, where the following condition is satisfied:
\begin{equation*}
  \rho_{\,v}\ -\ \rho_{\,a}\ \geq\ \varpi \cdot\bigl(\rho_{\,0}\ -\ \rho_{\,a}\bigr)\,,
\end{equation*}
where $\rho_{\,v}$ is the visible density and $\varpi$ is a constant whose approximate value belongs to the segment $[\,0.01,\, 0.1\,]\ \ni\ \varpi\,$. For the sake of illustration, we depicted this visible interfaces in Figure~\ref{fig:anal} with solid lines, while the exact analytical solutions were shown with dashed lines. In the preparation of the visible interface we assumed that the fluid density varies linearly inside the turbulent layer from $\rho_{\,a}$ to $\rho_{\,0}\,$. In other words, the virtual visible interface $y\ =\ y_{\,v}$ is determined by the following relation:
\begin{equation*}
  \rho_{\,v}\ -\ \rho_{\,a}\ =\ \bigl(\rho_{\,0}\ -\ \rho_{\,a})\cdot\Bigl(1\ -\ \frac{y_{\,v}}{h}\Bigr)\ =\ \varpi\cdot\bigl(\rho_{\,0}\ -\ \rho_{\,a}\bigr)\,.
\end{equation*}
For the sake of convenience, the last condition can be equivalently reformulated in terms of the buoyancy variable:
\begin{equation*}
  y_{\,v}\,(x)\ =\ \begin{dcases}
    \ h\,(x)\cdot\Bigl(1\ -\ \varpi\;\frac{b_{\,0}}{b\,(x)}\Bigr)\,,& b\,(x)\ \geq\ \varpi\,b_{\,0}\,, \\
    \ 0\,,& \mbox{otherwise.}
  \end{dcases}
\end{equation*}
\end{remark}

%%% ----------------------------------------------------------------------- %%%

\section{Derivation of the speed-Froude relation}
\label{app:der}

From ``boundary'' conditions \eqref{eq:bc2}, \eqref{eq:bc3} we deduce the following system of two equations with respect to unknowns $u$ and $c\,$:
\begin{align*}
  \bigl(u\ -\ c\bigr)\ +\ \half\;m\ &=\ 0\,, \\
  \bigl(c\ -\ u\bigr)^{\,2}\ &=\ m\,.
\end{align*}
By solving this system we obtain the following solution ($u\ \neq\ 1$):
\begin{equation*}
  u\ =\ \third\;c\,, \qquad
  m\ =\ \fourninth\;c^{\,2}\,.
\end{equation*}
Furthermore, by definition we have
\begin{equation*}
  \Fr_{\,b}\ =\ \frac{w_{\,b}}{\sqrt{m_{\,b}}}\,.
\end{equation*}
Thus,
\begin{equation*}
  U_{\,b}^{\,3}\ =\ b_{\,0}\,\zeta_{\,b}\,w_{\,b}\ =\ m_{\,b}\,w_{\,b}\ =\ \Fr_{\,b}\,m_{\,b}^{3/2}\,.
\end{equation*}
Hence,
\begin{equation*}
  m_{\,b}^{\,1/2}\ =\ \Fr_{\,b}^{\,-1/3}\,U_{\,b}\,.
\end{equation*}
Taking into account the last results, the ``boundary'' condition \eqref{eq:bc1} becomes:
\begin{equation*}
  m\cdot\bigl(c\ -\ u)\ +\ U_{\,b}^{\,3}\cdot\Bigl(\frac{c}{w_{\,b}}\ -\ 1\Bigr)\ =\ 0\,.
\end{equation*}
Or equivalently we have:
\begin{equation*}
  m\,\bigl(c\ -\ u)\,U_{\,b}^{\,-3}\ +\ \bigl(c\cdot\Fr_{\,b}^{\,-1}\cdot m_{\,b}^{-1/2}\ -\ 1\bigr)\ =\ 0\,.
\end{equation*}
By rearranging the terms in the last equation we obtain:
\begin{equation*}
  \frac{8}{27}\;\frac{c^{\,3}}{U_{\,b}^{\,3}}\ +\ \Fr_{\,b}^{\,-2/3}\;\frac{c}{U_{\,b}}\ -\ 1\ =\ 0\,.
\end{equation*}
By introducing the dimensionless velocity $C\ \eqdef\ \frac{c}{U_{\,b}}$ we obtain the required Equation~\eqref{eq:stst}.

%%% ----------------------------------------------------------------------- %%%

\section{Nomenclature}
\label{app:nom}

In the main text above we used the following notations (this list is not exhaustive):
\begin{description}
  \item[$\equiv\ $] equal identically
  \item[$\propto\ $] proportional
  \item[$\eqdef\ $] the left hand side is defined
  \item[$\defeq\ $] the right hand side is defined
  \item[$\apprle\ $] smaller than the approximate upper bound\footnotemark
  \item[$\apprge\ $] greater than the approximate lower bound\footnotemark[\value{footnote}]\footnotetext{We underline the fact that this bound is given approximatively.}
  \item[$\phi\ $] angle of the bottom slope
  \item[$\alpha\ $] local bottom slope, \ie~$\alpha\ =\ \tan\phi$
  \item[$h_{\,j}\ $] total depth of the layer $j$
  \item[$\rho_{\,j}\ $] fluid density in the layer $j$
  \item[$\zeta\ $] total depth of the bottom layer
  \item[$b_{\,j}\ $] buoyancy of the layer $j$
  \item[$u_{\,j}\ $] depth-averaged horizontal velocity of fluid particles in the layer $j$
  \item[$v\ $] vertical velocity component
  \item[$w\ $] horizontal velocity component in the bottom layer
  \item[$\mathrm{w}\ $] transversal velocity component (in 3D)
  \item[$p\ $] fluid pressure
  \item[$m_{\,j}\ \eqdef\ b_{\,j}\,h_{\,j}\ $] ``mass'' contained in the fluid column
  \item[$\M_{\,j}\ \eqdef\ m_{\,j}\,u_{\,j}\ $] mass flux in the layer $j$
  \item[$q\ $] depth-averaged turbulent mean square velocity in the mixing layer
  \item[$\xi\ $] characteristics speed (\ie eigenvalues of the \textsc{Jacobian} matrix of the hyperbolic fluxes)
  \item[$\lambda_{\,1,\,2}\ $] eigenvalues in the stability studies
  \item[$c\ $] dimensional velocity of the travelling wave
  \item[$C\ $] dimensionless wave velocity
  \item[$\Fr\ $] the dimensionless \textsc{Froude} number
  \item[$\ell,\,L\ $] length scales
  \item[$t\ $] time variable
  \item[$x\ $] ``horizontal'' coordinate along the bottom slope
  \item[$y\ $] ``vertical'' coordinate normal to the bottom
  \item[$d\,(x)\ $] the bathymetry (depth) function
  \item[$g\ $] gravity acceleration
  \item[$\tilde{g}\ $] reduced gravity acceleration
  \item[$\chi\ $] entrainment rate
  \item[$u_{\,\ast}\ $] friction velocity at the solid bottom
\end{description}

%%% ----------------------------------------------------------------------- %%%

%%% Bibliography
\addcontentsline{toc}{section}{References}
\bibliographystyle{abbrv}
\bibliography{biblio}
\bigskip

\end{document}